\documentclass{egpubl}
\usepackage{hpg25DL}

\WsPaper           

\usepackage[T1]{fontenc}
\usepackage{dfadobe}  

\usepackage{cite}  
\BibtexOrBiblatex
\electronicVersion
\PrintedOrElectronic
\ifpdf \usepackage[pdftex]{graphicx} \pdfcompresslevel=9
\else \usepackage[dvips]{graphicx} \fi

\usepackage{egweblnk} 

\usepackage{booktabs} 
\usepackage[table,xcdraw]{xcolor} 
\usepackage{svg}
\usepackage{algorithm}
\usepackage{algpseudocode}

\usepackage[normalem]{ulem}
\useunder{\uline}{\ul}{} 

\usepackage{hyperref}
\usepackage{xspace}
\usepackage{amsmath,amssymb} 
\usepackage{mathtools} 

\usepackage{subcaption}  

\usepackage{footnotehyper} 

\usepackage[utf8]{inputenc} 

\newcommand{\dataCa}{\mbox{\textsc{CA13}}\xspace}
\newcommand{\dataSwiss}{\mbox{\textsc{SwissS3d}}\xspace}
\newcommand{\dataBora}{\mbox{\textsc{Bund\_Bora}}\xspace}
\newcommand{\dataId}{\mbox{\textsc{ID15\_Bunds}}\xspace}
\newcommand{\dataNz}{\mbox{\textsc{Gisborne}}\xspace}
\newcommand{\dataNzA}{\mbox{\textsc{Gisborne\_A}}\xspace}
\newcommand{\dataNzB}{\mbox{\textsc{Gisborne\_B}}\xspace}
\newcommand{\dataNzC}{\mbox{\textsc{Gisborne\_C}}\xspace}

\newcommand{\name}{\mbox{\textsc{LidarScout}}\xspace}

\newcommand{\diff}[1]{#1}  

\newcommand\blfootnote[1]{%
  \begingroup
  \renewcommand\thefootnote{}\footnote{#1}%
  \addtocounter{footnote}{-1}%
  \endgroup
}

\title[LidarScout]%
      {LidarScout: Direct Out-of-Core Rendering of Massive Point Clouds}

\author[P. Erler \& L. Herzberger \& M. Wimmer \& M. Schütz]
{\parbox{\textwidth}{\centering P. Erler\thanks{perler@cg.tuwien.ac.at}$^{1}$\orcid{0000-0002-2790-9279} and L. Herzberger$^{1}$\orcid{0000-0002-9047-065X}  and M. Wimmer$^{1}$\orcid{0000-0002-9370-2663} and M. Schütz$^{1}$\orcid{0000-0002-8166-3089} 
        }
        \\
{\parbox{\textwidth}{\centering $^1$TU Wien, Austria 
       }
}
}

\begin{document}

\captionsetup{labelfont=bf,textfont=it}  

\teaser{
 \centering
 \includegraphics[width=\linewidth]{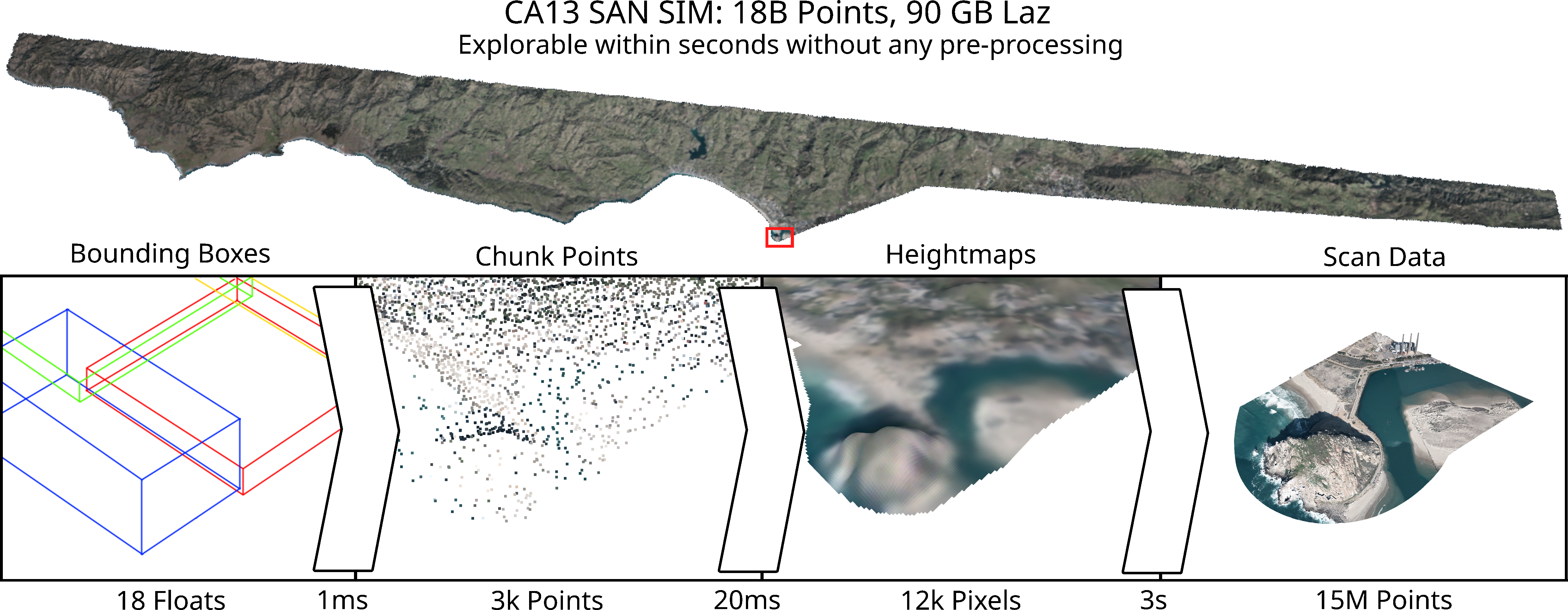}
  \caption{We present \name, a method to explore huge, compressed point clouds within seconds. The example shown here contains the Morro Bay area focusing on Morro Rock, rendered with three heightmaps of 640m x 640m.}
\label{fig:teaser}
}

\maketitle
\begin{abstract}
Large-scale terrain scans are the basis for many important tasks, such as topographic mapping, forestry, agriculture, and infrastructure planning. The resulting point cloud data sets are so massive in size that even basic tasks like viewing take hours to days of pre-processing in order to create level-of-detail structures that allow inspecting the data set in their entirety in real time.
In this paper, we propose a method that is capable of instantly visualizing massive country-sized scans with hundreds of billions of points. Upon opening the data set, we first load a sparse subsample of points and initialize an overview of the entire point cloud, immediately followed by a surface reconstruction process to generate higher-quality, hole-free heightmaps. As users start navigating towards a region of interest, we continue to prioritize the heightmap construction process to the user's viewpoint. Once a user zooms in closely, we load the full-resolution point cloud data for that region and update the corresponding height map textures with the full-resolution data. As users navigate elsewhere, full-resolution point data that is no longer needed is unloaded, but the updated heightmap textures are retained as a form of medium level of detail. 
Overall, our method constitutes a form of direct out-of-core rendering for massive point cloud data sets (terabytes, compressed) that requires no preprocessing and no additional disk space. 

Source code, executable, pre-trained model, and dataset are available at: \\\url{https://github.com/cg-tuwien/lidarscout}
\\


\end{abstract}  


\blfootnote{This work was published at HPG 2025: \url{https://diglib.eg.org/items/b044b2fe-88c1-4fe4-9ed2-2424fb2ed036}}

\section{Introduction}

Many fields require huge terrain scans, including archeology, infrastructure, bathymetry, agriculture, forestry, flood and landslide prediction, geology, climate research, and many more.
Improvements in laser scanners and frequent scanning operations (e.g., 3DEP~\shortcite{3DEPLAS} and AHN~\shortcite{AHN5}) result in country-wide data sets comprising hundreds of billions to trillions of points. These are typically stored in a compressed format (LAZ) and can amount to terabytes. Visualizing these data sets requires out-of-core level-of-detail (LOD) structures that allow loading only those tiny subsets necessary for a given viewpoint. However, generating these structures takes hours to days of preprocessing. For example, AHN2, a point cloud of the entire Netherlands, comprises 640 billion points, and constructing an LOD structure took Martinez-Rubi~\shortcite{MassivePotreeConverter} 15 days of processing.

With these huge amounts of data, tasks like viewing become non-trivial. Such supposedly simple tasks include having a quick overview, searching for obvious problems like outliers and noise, finding the relevant files for a specific region, and transferring the data. 
Compression can reduce transfer and storage problems, but makes other tasks even slower.
With \name, we reduce the time it takes to visualize massive data sets from days down to seconds. After a user drops the data set into the application, we quickly read all tiles' bounding boxes as the only global operation. Afterward, we efficiently pick a sparse subsample of the compressed point cloud, generate rough heightmaps, refine them with a small neural network, and render them with a CUDA-based software rasterizer, all prioritizing the user's current viewpoint. When zooming in further, we stream the high-resolution scan data and update the hightmaps.

Our main contributions are:
\begin{enumerate}
    \item An interactive point cloud viewer for massive terrain scans that requires no pre-processing and no additional disk space.
    \item Efficient extraction of a sparse subsample from compressed point clouds (LAZ).
    \item A method to predict high-quality heightmaps from this sparse subsample.
\end{enumerate}


\section{Related Work}

Related work includes fast point-based rendering approaches, particularly of massive data sets, and surface reconstruction, especially those related to constructing heightmaps from sparse point samples. We also briefly explore neural rendering methods, as several point-based neural methods reconstruct high-quality images from sparse point samples, a problem similar to constructing heightmaps from sparse subsamples.

\paragraph*{Surface Reconstruction}

Surface reconstruction aims to recover the underlying object that a point cloud was sampled from.
Most surface reconstruction methods work in full 3D to generate a mesh or distance field. Only a few works target 2.5D and directly output heightmaps.

Being one of the first 3D reconstruction methods, BallPivot~\cite{bernardini1999ball} runs from point to point in a point cloud, connecting them with edges. 
Screened Poisson Surface Reconstruction~\cite{kazhdan2013screened} is likely the most popular non-data-driven reconstruction method to date, despite requiring normals for fitting an indicator field to the points. BallMerge~\cite{parakkat2024ballmerge} uses Voronoi balls to recognize the inside and outside of large scans.
PPSurf~\cite{erler2024ppsurf} is a recent data-driven method predicting a signed-distance field from unoriented points. Many 3D reconstruction methods aim at single solids and tend to fail in open scenes.


Reconstructing heightmaps from point clouds has seen little work in recent years. However, the field of depthmap estimation is very active, and many results can be adapted to heightmaps.
Moving Least Squares~\cite{lancaster1981mls_surfaces} interpolates a smooth surface from scattered data points. The now classic approach for heightmap reconstruction is to generate a Delaunay triangulation and interpolate it linearly. This is done, e.g., in Las2Dem of the popular LAStools~\shortcite{rapidlassoGeneratingSpikeFree} suite, which was used for \dataSwiss~\shortcite{swissSURFACE3Draster}, for example. Las2Dem also deals with multiple laser returns falling into the same texel. However, triangulation approaches suffer from the fundamental problem of interpolating within slender triangles.
Closely related, most 2.5D reconstruction methods that work on depthmaps are in the context of single-view 3D reconstruction. Recent survey papers~\cite{rajapaksha2024deep, masoumian2022monocular} describe the advances of this field from depth-cue-based methods via machine learning and hand-crafted features to deep learning. 

Overall, surface reconstruction from point clouds has seen significant advances in recent years, especially on the data-driven side. However, heightmap reconstruction for aerial LIDAR scans has been neglected.



% ...

\paragraph*{Aerial LIDAR Storage}

The LAS and LAZ formats are specifically targeted towards aerial laser scanning and thus the two most commonly used formats for massive, country-wide aerial point cloud scans. LAZ is a compressed form of LAS that specifically takes advantage of common patterns in point clouds for efficient and lossless compression~\cite{LASzip}. For example, LAZ predicts the next position of a point based on the differences of the previous five points and then entropy encodes the difference between prediction and true position, which takes advantage of the fact that laser scanners observe points in a line-wise and thus predictable fashion.

\paragraph*{Point-Based Rendering}

Levoy and Whitted~\shortcite{PointAsDisplayPrimitive} proposed points as a meta-primitive that all other surface primitives can be converted to, or to be directly generated from procedural functions. Since then, point cloud rendering has become a widely popular field due to the vast amount of point samples generated by laser scanners, and the resulting need for higher performance and better quality~\cite{kobbelt2004survey}. Surfels~\cite{pfister2000surfels} and EWA-Splatting~\cite{zwicker2002ewa} introduced high-quality rendering approaches for point-based primitives. Botsch et al.~\shortcite{botsch2005high} propose an efficient GPU-based implementation for high-quality splatting. Günther et al.~\shortcite{Gnther2013AGP} and Schütz et al.~\shortcite{SCHUETZ-2022-PCC} improve the rendering performance via compute-based solutions that are faster than the triangle-oriented hardware pipeline. \diff{Schütz et al.~\cite{SMOW20} introduce a progressive rendering approach that renders random subsets each frame and converges towards the true result over the coarse of several frames.} 3D Gaussian Splatting~\cite{kerbl3Dgaussians} proposes 3-dimensional Gaussians as a geometric primitive for 3D reconstruction and novel-view-synthesis. While some of the referenced works deal with point-based geometry that comprises position as well as orientation and size/scale, our method is targeted toward point clouds from aerial LIDAR scans, whose geometry is solely described by positions.

\paragraph*{Point-Based Levels of Detail}

Rendering massive data sets requires LOD structures that reduce memory usage and increase performance. QSplat~\cite{QSplat} proposes a bounding-sphere hierarchy as a means to render large splat models. Sequential Point Trees~\cite{Dachsbacher2003} introduces a similar hierarchy but sequentializes it into an array that allows efficient rendering on the GPU by invoking a draw call for a subset of the array, replacing fine-grained hierarchy traversal by a draw call to a batch of data. Instant Points~\cite{InstantPoints} suggests a nested octree that allows for view-dependent LOD. Wand et al.~\shortcite{Wand2008} and Modifiable Nested Octree (MNO)~\cite{scheiblauer2011} propose modifiable structures that enable efficient selection and deletion on large point data sets. Potree~\cite{SCHUETZ-2020-MPC} and Lidarserv~\shortcite{RealTimeIndexingBormann} improve the LOD construction performance of MNOs. \diff{Lidarserv and SimLOD~\cite{schutz2024simlod} both propose incremental LOD construction algorithms that build and display the LOD structure while additional points are loaded. The former focuses on live-capture of point clouds, while the latter focuses on GPU-accelerated LOD construction that can build the structure as fast as points can be streamed from disk. }

The largest LOD construction study for point clouds that we are aware of was made by Martinez-Rubi et al. -- 640 billion points converted to an MNO in 15 days~\cite{MassivePotreeConverter}.


\paragraph*{Neural Point Cloud Rendering}

Neural rendering uses a neural network to synthesize images for given parameters, rather than generating and rasterizing geometry. Tewari et al.~\shortcite{tewari2020state} give an overview in their STAR. Neural Point Cloud Rendering was just a side-note in 2020, with Neural Point-Based Graphics~\cite{aliev2020neural} being the only mention. First, they compute feature vectors for the given points. Then, they rasterize them as high-dimensional points in multiple resolutions. Finally, they feed these raw images into a U-Net~\cite{ronneberger2015unet}, which outputs a rendering. Since then, many methods have been proposed~\cite{kopanas2021point, nguyen2021rgbd, wang2021ibrnet, ruckert2022adop, rakhimov2022npbg++, you2023learning, harrer2023inovis, franke2024trips, hahlbohm2024inpc}, improving various aspects of the approach. These methods share one drawback: they are made for relatively dense point clouds, typically produced by photogrammetry, and many require camera poses, which are not available in our application.

\section{Method}
\label{sec:method}

\name consists of five stages: Quickly loading a sparse subsample of the entire point cloud; generating heightmaps with textures; refining them for the user's viewpoint; loading full-res data in close-up viewpoints and updating heightmaps with full-res data to retain a medium level of detail; and rendering them with a CUDA software rasterizer.


\subsection{Data and Data Structures}
\label{sec:method_data}

Massive aerial LIDAR data sets are typically distributed using the compressed LAZ format (e.g., OpenTopography~\shortcite{CA13_SAN_SIM}, AHN5~\shortcite{AHN5}, 3DEP\shortcite{3DEPLAS}, etc.). For this paper's evaluation, we selected three point clouds from the USA, one from New Zealand, and one from Switzerland, as shown in Table~\ref{tab:datasets}. The sizes range from 1.6 billion points (ID15\_BUNDS) to 262 billion points (Gisborne+Addendum). The full Switzerland data set would exceed Gisborne \diff{(estimated 18 TB Zipped LAS)}, but due to lack of storage we selected a subset of 18.7 GB. In the context of aerial LIDAR scans, a point's geometry is solely defined by its position, unlike other point-based primitives such as Surfels or Gaussian Splats. In many cases, they also lack colors, which is why we include the data set of Switzerland as a representative example. 

\begin{figure*}[t]
    \centering
    \begin{subfigure}{0.24\linewidth}
        \includegraphics[trim=0 145 0 170,clip,width=\textwidth]{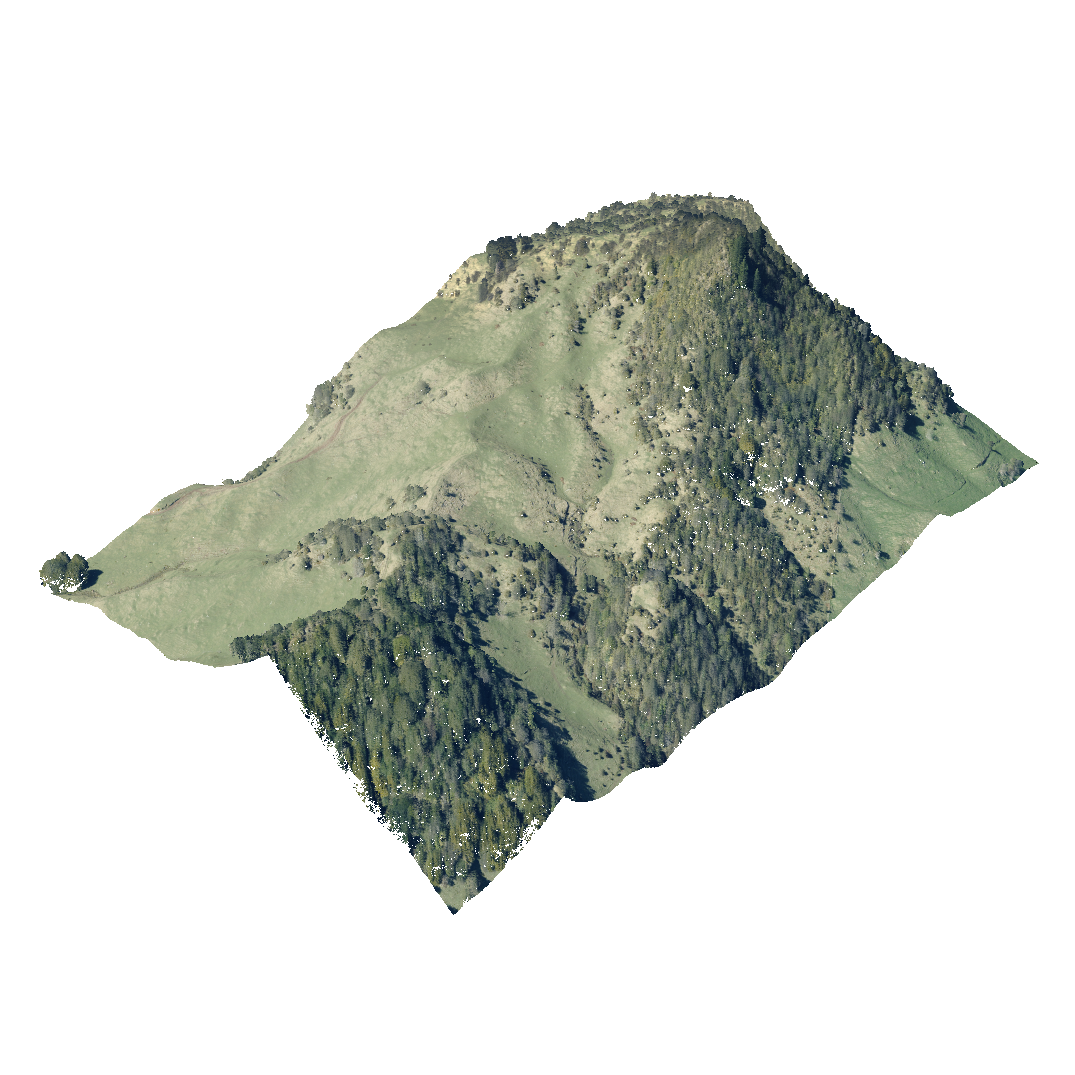}
        \caption{Tile}
        \label{fig:first}
    \end{subfigure}
    \hfill
    \begin{subfigure}{0.24\linewidth}
        \includegraphics[trim=0 145 0 170,clip,width=\textwidth]{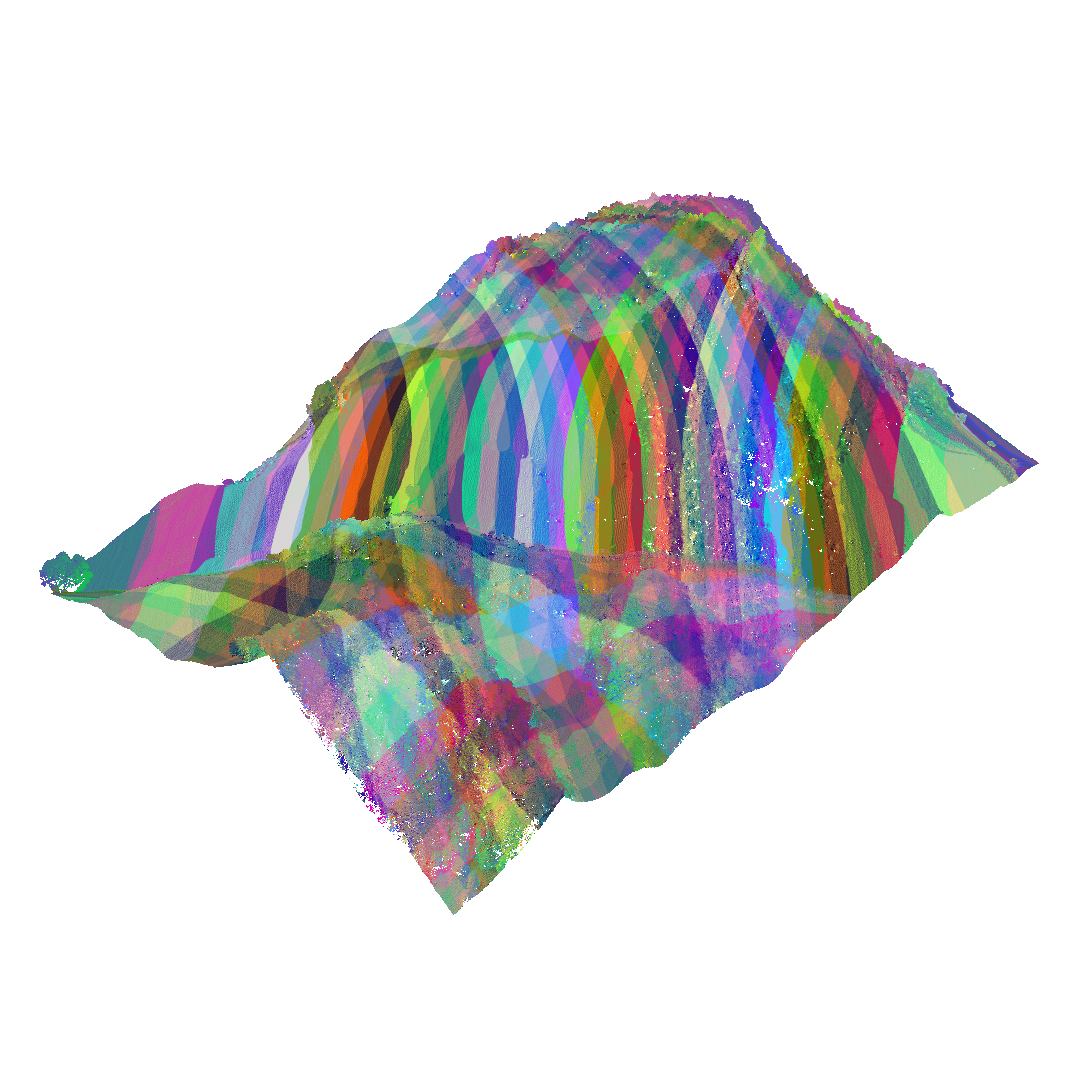}
        \caption{Chunks}
        \label{fig:second}
    \end{subfigure}
    \hfill
    \begin{subfigure}{0.24\linewidth}
        \includegraphics[trim=0 145 0 170,clip,width=\textwidth]{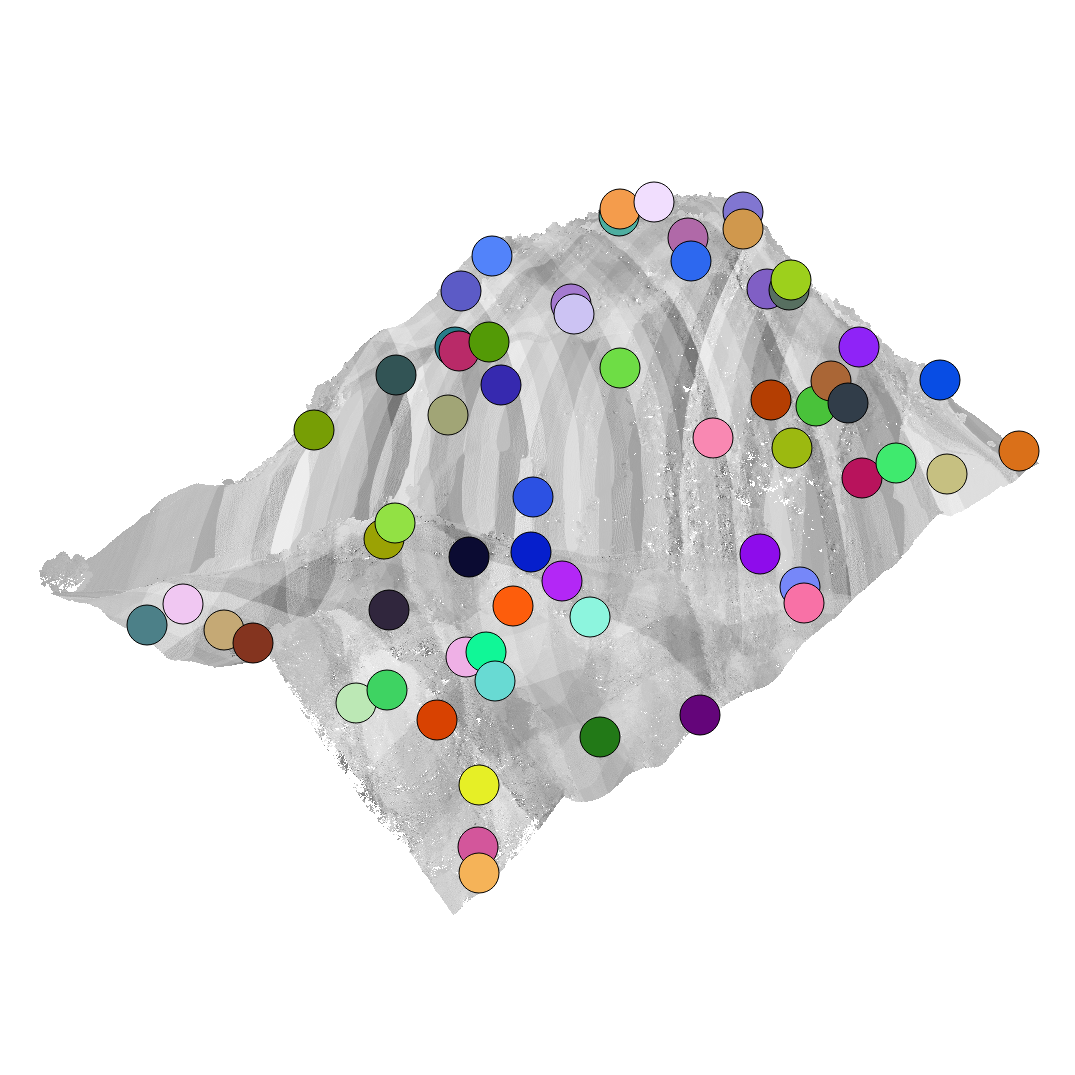}
        \caption{Chunkpoints}
        \label{fig:third}
    \end{subfigure}
    \hfill
    \begin{subfigure}{0.24\linewidth}
        \includegraphics[trim=0 50 0 10,clip,width=\textwidth]{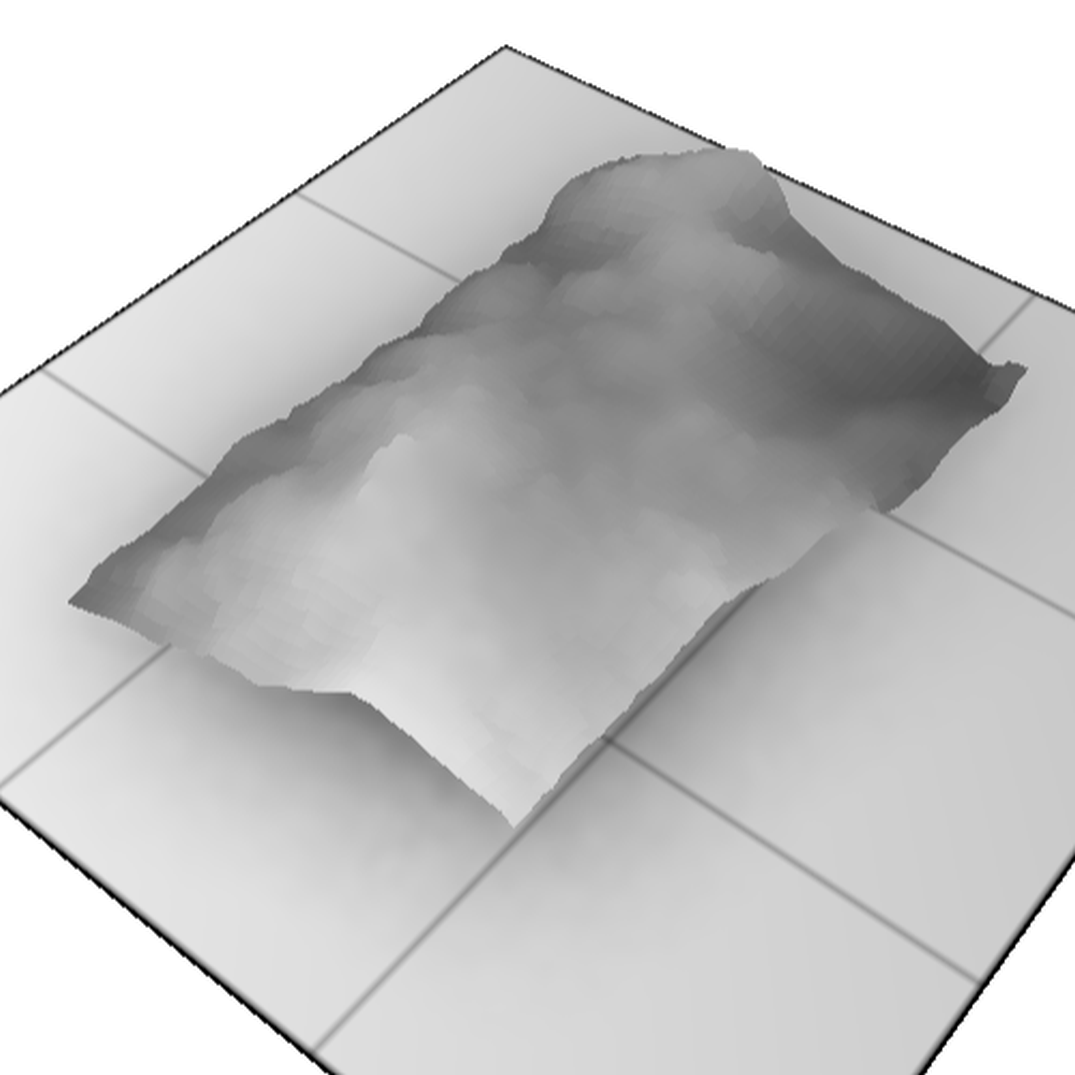}
        \caption{Patches}
        \label{fig:fourth}
    \end{subfigure}
    \hfill
    \caption{Massive LIDAR data is stored in rectangular tiles. Tiles store data in compressed chunks of 50k points. Points in a tile are typically stored by timestamp, in this case indicating circular scanning patterns. Chunk points refer to the uncompressed first point of each chunk. Patches denote 640x640 meter (10m/pixel) heightmaps, created from the rapidly loaded chunk points.}
    \label{fig:tiles_chunks_patches}
\end{figure*}

In this paper, we will regularly refer to tiles, chunks, chunk points, and patches, as illustrated in Figure~\ref{fig:tiles_chunks_patches}. \textbf{Tiles} correspond to individual LAZ files. Massive LIDAR data sets are almost always organized in such rectangular tiles, typically covering several hundreds to a thousand meters, and storing several millions of points. \textbf{Chunks} are an important concept in the LAZ compression algorithm. LAZ compresses points in chunks of 50\,000 points.
The first point in each chunk is uncompressed, and subsequent points must be decoded sequentially. Multiple chunks may be decompressed in parallel. \textbf{Chunk Points} refer to the first uncompressed point in each chunk. These are important in our approach since they are the only ones we can quickly access without the need to set up an expensive arithmetic decoder. \textbf{Patches} are quadratic 640 meter x 640 meter regions for which we reconstruct a 64 x 64 pixel heightmap from the surrounding sparse chunk points.

\begin{table}[!ht]
\centering
\resizebox{\columnwidth}{!}{%
\begin{tabular}{|m{0.18\textwidth} m{0.32\textwidth}|}
\hline
\textbf{Data set} & \textbf{Image} \\
\hline
\underline{\textbf{\dataCa}} \newline 
\href{https://portal.opentopography.org/raster?opentopoID=OTSDEM.032013.26910.2}{OpenTopography} \newline 
17.7 B Points \newline 
80 GB LAZ \newline 
20 $Points/m^2$ \newline 
LIDAR (linear) \newline
Screenshot: 56M points& \includegraphics[trim=0 50 0 20,clip,width=0.33\textwidth]{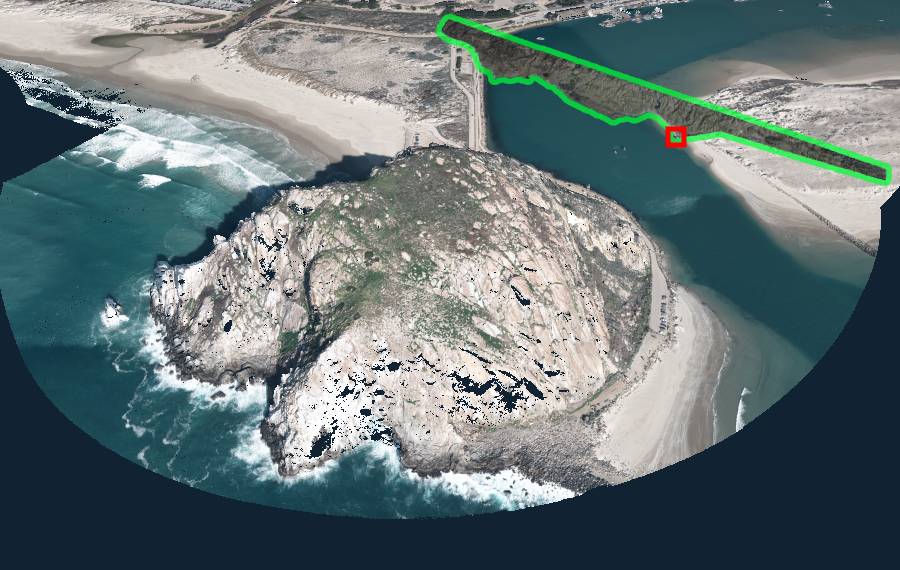} \\
\hline
\underline{\textbf{\dataSwiss}} \newline 
\href{https://www.swisstopo.admin.ch/de/hoehenmodell-swisssurface3d}{Swisstopo} \newline 
\diff{700 M Points (of 714 B)} \newline 
\diff{18.7 GB LAS (of 18 TB)} \newline 
17 $Points/m^2$ \newline 
No colors \newline
LIDAR (linear) & \includegraphics[trim=0 70 0 170,clip,width=0.33\textwidth]{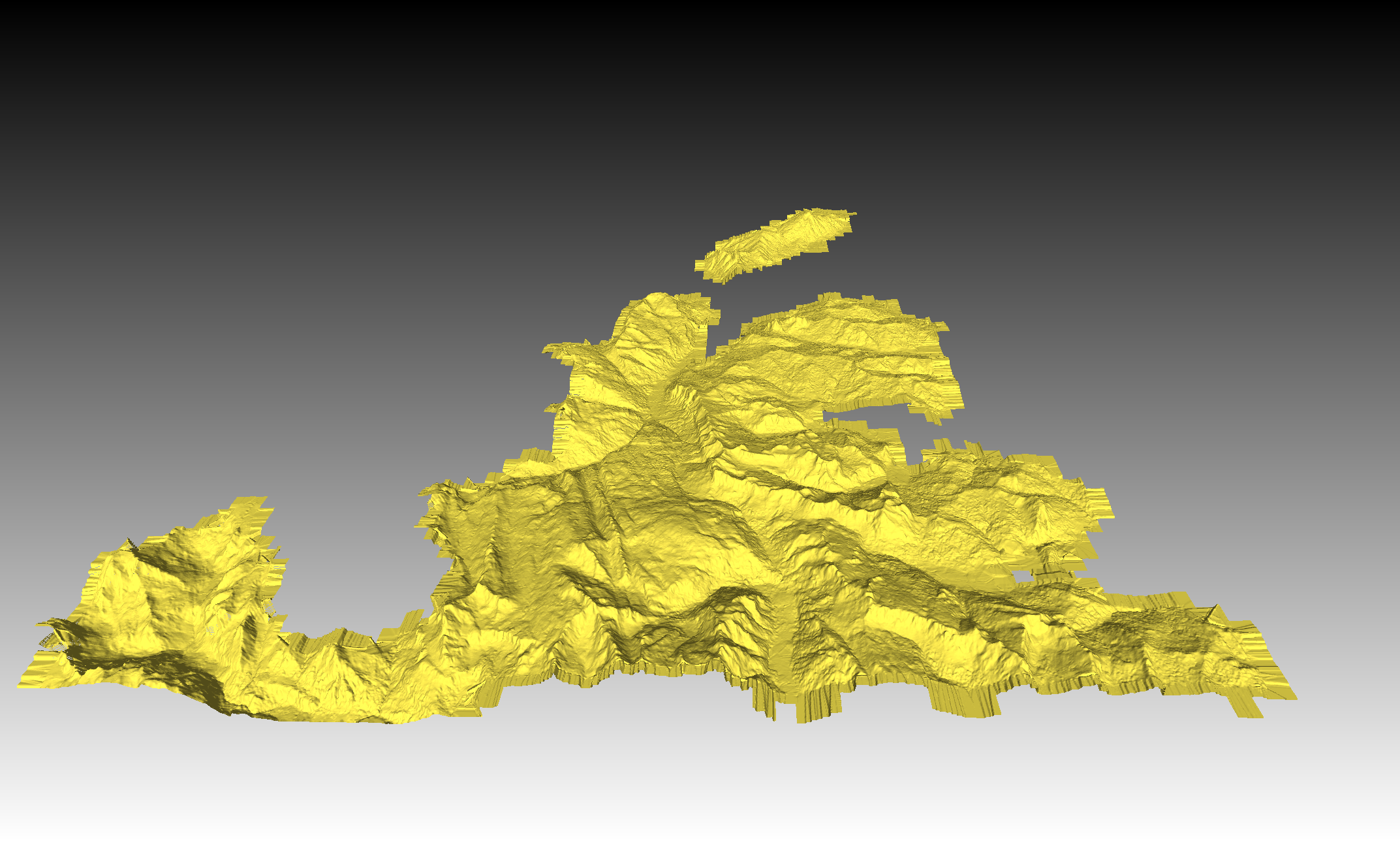} \\
\hline
\underline{\textbf{\dataBora}} \newline 
\href{https://portal.opentopography.org/datasetMetadata?otCollectionID=OT.052019.6341.1}{OpenTopography} \newline 
6.5 B Points \newline 
40 GB LAZ \newline 
574.22 $Points/m^2$ \newline 
Photogrammetry & \includegraphics[width=0.33\textwidth]{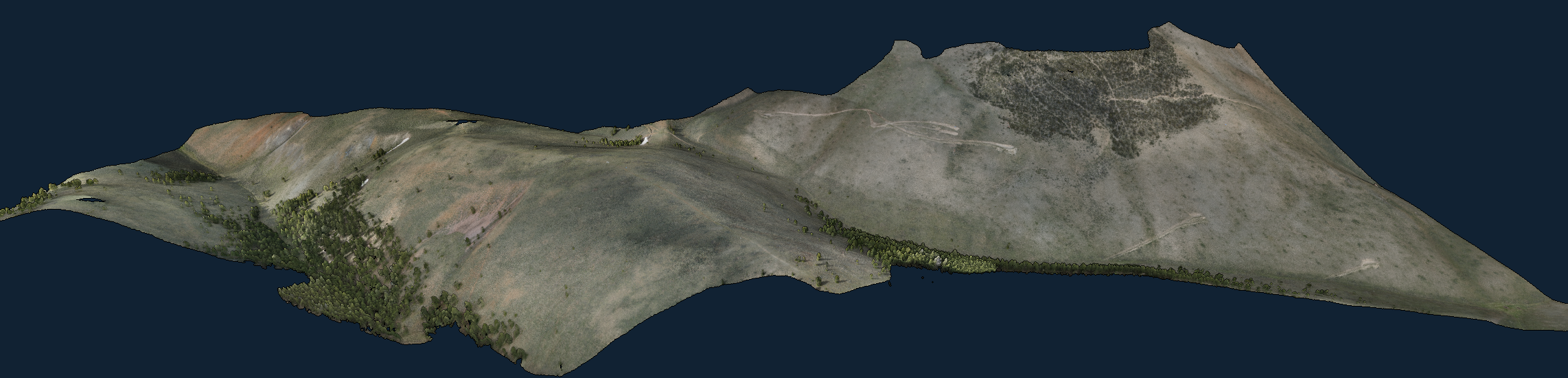} \\
\hline
\underline{\textbf{\dataId}} \newline 
\href{https://portal.opentopography.org/datasetMetadata?otCollectionID=OT.112020.6341.1}{OpenTopography} \newline 
1.6 B Points \newline 
14.8 GB LAZ \newline 
387.24 $Points/m^2$ \newline 
Photogrammetry & \includegraphics[trim=0 180 0 10,clip,width=0.33\textwidth]{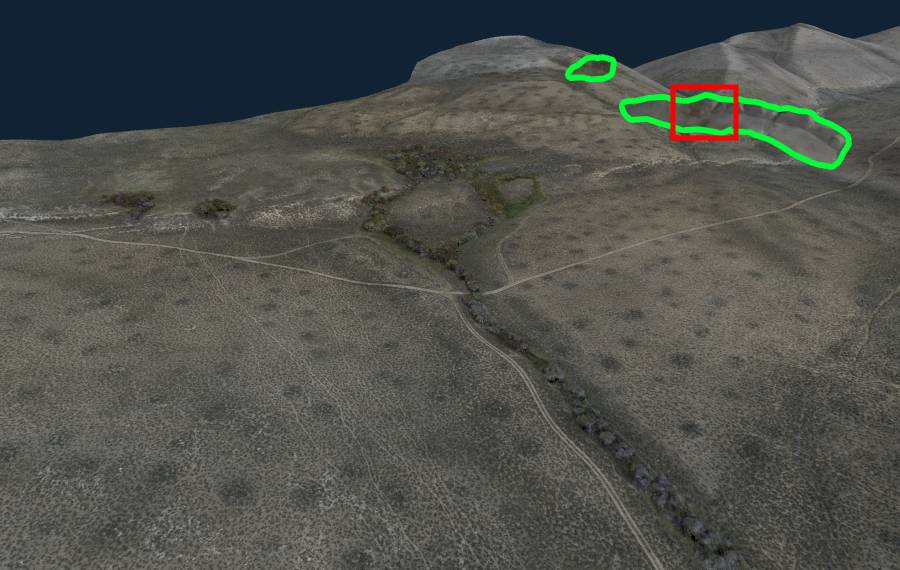} \\
\hline
\underline{\textbf{\dataNz}}+Addendum \newline 
\href{https://portal.opentopography.org/datasetMetadata?otCollectionID=OT.022024.2193.1}{OpenTopography} \newline 
95 + 167 B points \newline 
850 + 1\,550 GB LAZ \newline 
30.25 $Points/m^2$ \newline 
LIDAR (circular) \newline
Screenshot: 248M points & \includegraphics[trim=0 150 0 10,clip,width=0.33\textwidth]{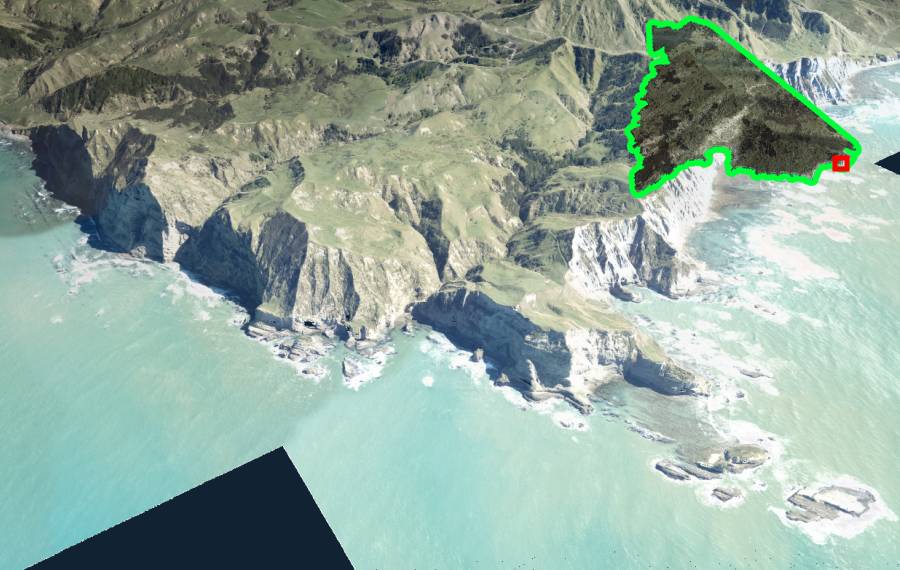} \\
\hline
\end{tabular}
}
\caption{Data sets used for \name. The table shows close-up screenshots, the entire map with green outline, and the close-up's area marked as red box. }
\label{tab:datasets}
\end{table}



\subsection{Rapidly Creating an Overview for Billions of Points}
\label{sec:method_data_loading}

In the first stage, we initialize an overview of the entire data set with a sparse subsample, which poses two challenges: File I/O is optimized for loading sequential data instead of random subsamples. Also, massive data sets are typically compressed sequentially, which further limits our ability to access random sparse subsets. 

On modern SSDs, the first challenge is addressed by investigating their 4 kB random access performance. Similar to accessing RAM~\cite{drepper2007every} or GPU memory~\cite{GlobalMemoryAccess}, SSDs are also optimized for coalesced access to a range of bytes rather than a few individual bytes. In case of SSDs, access is typically grouped into sectors of 4 kB, i.e., fetching a single point from disk is about as fast as accessing all points in a sector. However, only the first is uncompressed and thus easily accessible. Modern SSDs are capable of reading about 1 million random 4 kB sectors per second, so we are theoretically able to load a subsample of 1 M points of an arbitrarily large data set -- an arguably sufficiently large subset for an overview viewpoint on a 2-megapixel monitor. For example, the model shown in Figure~\ref{fig:teaser} depicts the heightmap model that was constructed from that data set's $\frac{18 B}{ 50 k}  = 360 k$ chunk points. 

The second challenge -- the industry standard LAZ compression format for massive aerial LIDAR data -- puts a limit on the way we can load the initial sparse sample. LAZ uses arithmetic coding to sequentially compress points one after the other, and in turn we also need to decompress them sequentially~\cite{LASzip}. Fortunately, points are compressed in chunks of typically 50 k points, and the first point in each chunk remains uncompressed, thus we are able to quickly load a sparse subsample made of every 50\,000th point. Since LAZ is variable-rate compressed, byte locations of each uncompressed chunk point must be obtained by first reading the file's chunk table. 

After all chunk points are loaded, we have a sparse subsample of the entire data set that is sufficiently dense in an overview perspective. When zooming in, holes between points will appear. Since massive aerial LIDAR data sets constitute 2.5D data until one zooms in closely, we propose to fill these holes by constructing high-quality heightmaps from the sparse set of chunk points. We divide the entire map into patches, covering 640x640 meters each, and for each patch, we construct a 64x64 pixel textured heightmap.

\subsection{Interpolated Heightmaps}
\label{sec:method_raw_hm}

\begin{algorithm}[ht!]
\caption{\diff{Patch-Space Chunk Points to Heightmaps}}
\label{alg:triangulation_interpolation}
\begin{algorithmic}[1]
\Require Chunk points $P = \{p_i\}$ (with $p_i \in [-1, 1]^2$), height values $h = \{h_i\}$, (optional) color $rgb = \{c_i\}$, grid resolution $res$
\Ensure Heightmaps $\texttt{hm}_{\text{nn}}$, $\texttt{hm}_{\text{lin}}$, mask $\texttt{face\_map}$

\State \textit{//~~~Initialization}
\State $N \gets res^2$
\State $K \gets \text{KDTree}(P)$
\State \textbf{Add} corner padding points (with NaN values) to $P$, $h$ and $rgb$
\State $\mathcal{T} \gets \Call{DelaunayTriangulate}{P}$
\State $G \gets$ generate regular grid of $res \times res$ points over $[-1, 1]^2$
\State Initialize mask array $\texttt{face\_map}[1..N]\leftarrow -2$
\State \textbf{Allocate} $\texttt{hm}_{\text{nn}}[1..N]$ and $\texttt{hm}_{\text{lin}}[1..N]$

\State \textit{//~~~FloodFill from Triangle Centroids}
\ForAll{triangles $T_i$ in $\mathcal{T}$}
    \State $c_i \gets$ centroid of $T_i$
    \State $g_{\text{id}} \gets$ index of $G$-grid point closest to $c_i$
    \State \Call{FloodFill}{$g_{\text{id}}$, $i$, $\texttt{face\_map}$, $\mathcal{T}$, $G$, $res$}
\EndFor

\State \textit{//~~~FloodFill from Disconnected Rests}
\ForAll{triangles $T_i$ in $\mathcal{T}$}
    \State $B_i \gets$ bounding box of $T_i$ intersected with $[-1,1]^2$
    \ForAll{grid points $g_j$ in $B_i$}
        \If{$\texttt{face\_map}[j] = -2$ \textbf{and} $g_j \in T_i$}
            \State \Call{FloodFill}{$j$, $i$, $\texttt{face\_map}$, $\mathcal{T}$, $G$, $res$}
        \EndIf
    \EndFor
\EndFor

\State \textit{//~~~Remove Face IDs of Padding Triangles}
\ForAll{$j = 1 \ldots N$ with $\texttt{face\_map}[j] \ge 0$}
    \State $i \gets \texttt{face\_map}[j]$
    \If{any vertex of triangle $\mathcal{T}[i]$ is a padding vertex}
        \State $\texttt{face\_map}[j] \gets -1$
    \EndIf
\EndFor

\State \textit{//~~~Linear Interpolation in Convex Hull}
\ForAll{$j$ with $\texttt{face\_map}[j] \ge 0$}
    \State $i \gets \texttt{face\_map}[j]$
    \State $T \gets \mathcal{T}[i]$
    \State $\text{bary} \gets$ barycentric coordinates of $G[j]$ in $T$
    \State $\texttt{hm}_{\text{lin}}[j] \gets$ interpolate $h$ at triangle vertices of $T$ via $\text{bary}$
    \State Optionally: color $rgb$ via barycentric interpolation
\EndFor

\State \textit{//~~~Linear Interpolation outside Convex Hull}
\ForAll{$j$ with $\texttt{face\_map}[j] = -1$}
    \State $\texttt{hm}_{\text{lin}}[j] \gets$ nearest neighbor interpolation on $h$ via $K$ at $G[j]$
    \State Optionally: color $rgb$ via nearest-neighbor interpolation
\EndFor

\State \textit{//~~~NN Interpolation}
\ForAll{$j = 1\ldots N$}
    \State $\texttt{hm}_{\text{nn}}[j] \gets$ nearest neighbor interpolation on $h$ via $K$ at $G[j]$
\EndFor

\State \textit{//~~~Finish Up}
\State Remove padding points from $P$, $h$, $rgb$ if needed\\
\Return $\texttt{hm}_{\text{nn}}$, $\texttt{hm}_{\text{lin}}$, $\texttt{face\_map}$
\end{algorithmic}
\end{algorithm}


The goal of this part is to convert a region of chunk points to rough, textured heightmaps, which we will later refine with a neural network. For any patch of 64x64 pixels (10m/pixel) we first construct two 96x96 pixel heightmaps using nearest-neighbor (NN) and linear interpolation of the triangulated chunk points. The additional padding is applied to avoid seams between adjacent learned heightmaps. 

\diff{We receive $\mathcal{P}_{ms}$, the relevant chunk points for the current patch, from a grid-based data structure. We transform these points from model space to patch space as follows:
$\mathcal{P} = (\mathbf{p} - \mathbf{c})/r, \mathbf{p} \in \mathcal{P}_{ms},$
where $\mathbf{c} \in \mathbb{R}^2$ is the 2D patch center and $r \in \mathbb{R}$ is the padded patch radius. 
The triangulation and interpolation for the heightmaps and textures work as described in Algorithm~\ref{alg:triangulation_interpolation}, omitting minor implementation details and optimizations for clarity. At the core, we fill a face map indicating which triangle of the triangulation covers which pixel. This is mostly done by performing Flood-Fill from the triangle centroids. Due to rasterization, some pixels of very slender triangles may be disconnected. We catch those in a second Flood-Fill step, started at every pixel inside the bounding box of the triangle. The Flood-Fill works in an 8-connected neighborhood. It fills pixels if they are inside a triangle by checking their barycentric coordinates. After collecting the triangle IDs, we interpolate linearly, again with barycentric coordinates. This is only possible inside the convex hull of the triangulation. Therefore, we fall back to NN interpolation on the outside using a KD-Tree of $\mathcal{P}$ for the necessary speed. The NN interpolation uses the same KD-Tree for all pixels.}

\subsection{Learned Heightmaps}
\label{sec:method_clean_hm}

The linearly interpolated patches could be used directly for rendering. However, discontinuities and interpolation across slender triangles reduce the visual quality. Therefore, we employ a small neural network that produces accurate and visually pleasing interpolations.


The input of our CNN consists of two 96x96 heightmaps and two 96x96x3 RGB textures, both linearly and NN interpolated. It outputs a 64x64 heightmap and a 64x64x3 RGB texture. The 16 additional texels in each direction give the network context information, so it can produce smooth patch seams. Redundancy in the outputs could smooth the seams further but is not necessary in practice. We batch inference calls to avoid kernel overhead and context switches, increasing efficiency.

Since we need to keep every step of our method local, the model must not depend on the global context. \diff{Therefore, we normalize the given heights to patch space with $z = (z_{ms} - c_z) / r$, where $z_{ms}$ is a height in model space (meters above sea level) and $c_z$ is the height of the patch center. This normalization also helps avoid numerical problems in the network. Since the patch centers are created from a grid on the $XY$-plane, we take $c_z$ from the linearly interpolated heightmap's center.} This means that the network cannot know how far above sea level a patch is, which makes it more general but also makes color estimation more difficult.

\begin{figure}[h]
    \centering
    \scriptsize
    \includegraphics[width=\linewidth]{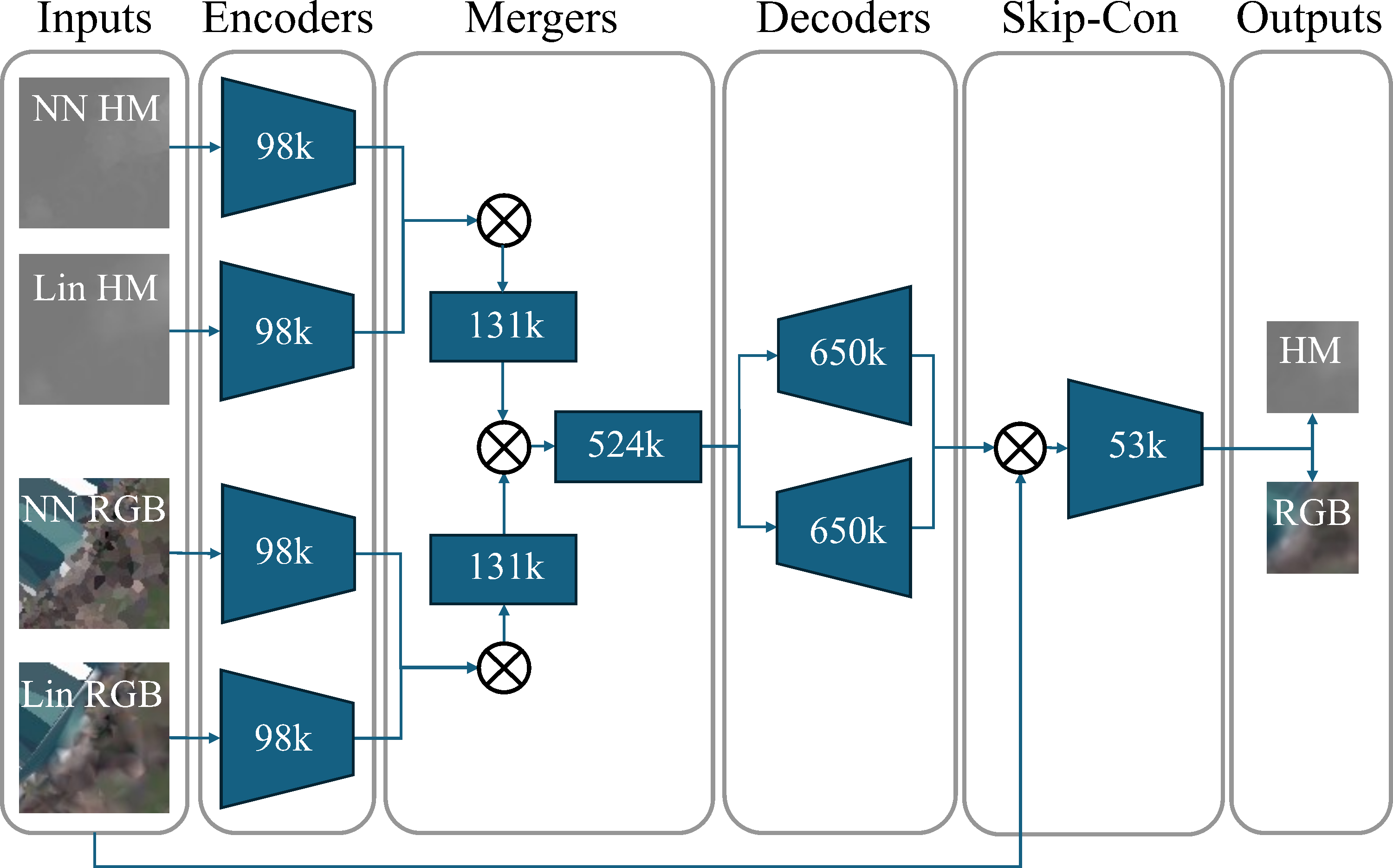}
    \caption{\name architecture. The network predicts clean heightmaps and textures from rough ones using a combination of encoders, fully-connected layers, and decoders. }
    \label{fig:architecture}
\end{figure}

\begin{figure*}[h]
    \centering
    \includegraphics[width=\linewidth]{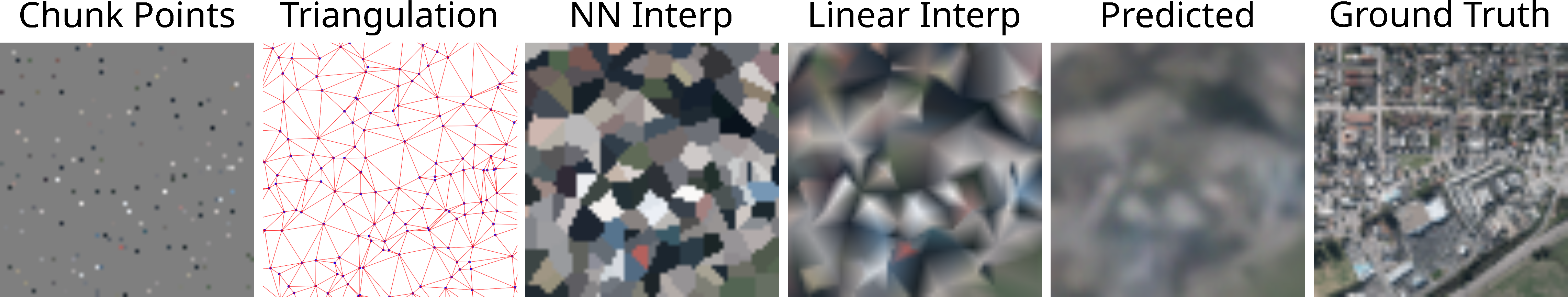}
    \caption{\diff{Example patch with inputs, processing steps, and network prediction.}}
    \label{fig:patch_example}
\end{figure*}

\diff{Our network architecture (see Section~\ref{fig:architecture}) is inspired by the U-Net~\cite{ronneberger2015unet}, which is also used in Neural Point-Based Graphics (NPBG)~\cite{aliev2020neural}. Unlike NPBG, we encode each input independently with several convolution layers, which increases the model size. However, since we work with dense inputs, we do not need gated convolutions, making our network smaller. Next, we merge the produced feature vector with separate, fully connected layers in two steps. Then, we decode the feature vector with two separate CNNs to heightmap and texture. In contrast to NPBG, we only have one skip connection that concatenates the original inputs and the decoder outputs along their channel dimension. Another CNN reduces this tensor again to the final number of channels. Finally, we take the center 64x64 region for further usage in rendering. In total, the network has 2.5M learnable parameters, 500k more than NPBG. Figure~\ref{fig:patch_example} shows one example patch with its chunk points, triangulation, interpolations, prediction, and ground truth.} The GT cannot be reached from the available sparse chunk points. Compared to linear interpolation, our network prediction is smoother, usually more accurate, and provides better transitions for our implicit level of detail. Note the red outlier at the middle bottom, which interpolation and simple smoothing preserve, but our CNN manages to ignore.

We train \name only on \dataCa and \dataSwiss, while we evaluate it on all the data sets described in Section~\ref{sec:method_data} to show its generalizability. For each data set, we generate our ground-truth data for training and evaluation from these scans. We choose the \diff{patch centers} randomly from the entire point cloud. 7000 of them are for training, 3000 for evaluation. For \dataCa and \dataSwiss, we split the \diff{patch centers} by the 70-percentile of their x coordinates into train and test sets. For each \diff{patch center}, we sample a heightmap and a texture by taking the mean of all points that fall onto a texel, which also removes most of the original scanning noise. To simulate the chunk points of LAZ, we randomly select a 50\,000th of all points.


We train our network with MSE loss, supervised by the ground-truth patches. We set loss elements corresponding to NaN elements in the GT (gaps in the original scans) to zero. The heightmap and texture losses are clipped to $(0.0\dots1.0)$ and averaged. We tried loss functions that put more weight on tile seams or gradients, but they did not make a noticeable difference. L1 loss and SSIM produce very similar, smoothed results, and LPIPS loss creates stripe artifacts. The bad results with LPIPS are likely due to a low number of top-down landscape images in their data sets and our images having a very low resolution.
Our optimizer is AdamW (lr = 0.0001, betas = (0.9, 0.999), eps = 1e-5, weight decay = 1e-2, amsgrad = False). With a step scheduler (gamma = 0.1, steps at 25 and 50 epochs), we train for 75 epochs. The training on \dataCa and \dataSwiss takes about 25 minutes. 
Training is done in Python using PyTorch, and the inference in C++/CUDA using LibTorch. 



\subsection{Full-Resolution Tiles}
\label{sec:method_full_res_tiles}

Upon navigating close to the surface, we additionally load and display full-resolution point data from tiles with a sufficiently large screen-space bounding box. In our test data sets, tiles typically hold about 1 to 50 million points. Close-range viewpoints such as in Figure~\ref{fig:screenshots}
may require loading about 50-300 million points, but nowadays, compute-based brute-force software rasterizers are capable of rendering up to two billion points in real time~\cite{SCHUETZ-2022-PCC}. Tiles that are no longer in focus are unloaded to free memory for other tiles. However, we update the heightmap textures to preserve some information for medium zoom levels.

\subsection{Rendering}
\label{sec:method_rendering}

We need to render chunk points for any patch whose heightmap is not yet ready, followed by rendering textured heightmaps, and eventually by rendering the full-resolution point cloud data upon zooming in close to a tile. Rendering of points and heightmaps was implemented in a CUDA-based software rasterizer. 

For points, we use the approach by Schütz et al.~\shortcite{SCHUETZ-2022-PCC}: We launch one thread-block comprising 256 threads for each chunk of 50k points. The threads iterate through all points, projecting them to screen and encoding their depth and color value into a 64 bit integer. We then use a 64 bit atomicMin to evaluate the point with the smallest depth value for each pixel. Afterward, a screen-space resolve pass extracts the color value from the least significant bits of each pixel's 64 bit depth and color value, and stores the result in an OpenGL texture for display. 

Heightmaps are rendered with a custom CUDA-based triangle-rasterizer: We invoke a cooperative kernel with 64 threads per block. Each block processes 32 triangles at a time, projects them to screen, and computes the screen-space bounding box. For any triangle that covers less than 1024 pixels, a single thread of the group iterates over the pixels, evaluates the barycentric coordinates. If they indicate that the pixel is inside the triangle, it draws the fragment using the same 64 bit atomic-min logic as the point rasterizer. If a triangle is larger than 1024 pixels, it is added to a queue. After the block finishes rendering the smaller triangles in one thread per triangle, it continues to render the large triangles utilizing all 64 threads for each triangle, one after the other. The thread blocks continue to loop until all triangles of all heightmaps are rendered.

\section{Results}
\label{sec:results}

We evaluate two use cases for our method and compare with several baselines: exploring large point clouds quickly and estimating the surface accurately from local subsamples. 

\subsection{Exploring Large Point Clouds}

\begin{table}[]
\caption{Time to construct an LOD structure in Potree vs. time to completely finish each stage in \name. 
After loading all tiles' metadata, loading the chunk points and the stages of heightmap construction operate progressively, so users can already navigate the data set before they are finished.
}
\small
\setlength\tabcolsep{4.5pt}
\begin{tabular}{@{}l|r|rrr@{}}
\toprule
Data set     & Potree         & \multicolumn{3}{c}{\name}        \\ 
             &                & Tiles   & Chunk Points  & Heightmaps     \\ \midrule
\dataCa      & 28m44s         & 0.02s   &  1.9s         & 23s \\
Gisborne     & 19h56m         & 0.10s   & 12.8s         & 159s \\
Gisborne+Add & -              & 0.19s   & 53.0s         & 701s \\ \bottomrule
\end{tabular}
\label{tab:potree_compare}
\end{table}

In order to display massive data sets that do not fit in GPU memory, state-of-the-art solutions require creating LOD structures in a preprocessing step. Potential solutions include Entwine~\shortcite{Entwine}, Potree~\shortcite{SCHUETZ-2020-MPC}, MassivePotreeConverter~\shortcite{MassivePotreeConverter}, and Lidarserv~\shortcite{RealTimeIndexingBormann}. We will study the performance of \name in comparison to Potree, the fastest of the related approaches. For rendering, we use this test system: RTX 4090; AMD Ryzen 9 7950X 16-Core; Crucial T700 4TB PCIe Gen5.

\subsubsection{Case Study: CA13 (17.7 billion points).}

As shown in Table~\ref{tab:potree_compare}, Potree takes 28m44s until LODs are constructed and the data set can be explored. \name takes 0.02s to load the metadata of 2336 tiles and another 1.9s  until all chunk points are loaded. Heightmap construction takes 23s  to finish, but construction prioritizes the current viewpoint so users do not need to wait to see meaningful results. 

\subsubsection{Case Study: Gisborne (95 billion points).} 

\name takes 12.8s  until a subsample of 1.9 million chunk points has been loaded in order to display an overview of the entire data set. Heightmaps are then constructed based on the user's current viewpoint. In comparison, users would traditionally have to wait 19h56m to construct an LOD structure before being able to explore the data set. The LOD structure that was constructed by Potree required 1.7TB of additional disk space.

\subsubsection{Case Study: Gisborne+Addendum (262 billion points)} 

We were not able to evaluate Potree's performance due to lack of additional disk space for the constructed LOD data. Extrapolating from Gisborne without Addendum, Potree would presumably require 2 days and 6 hours to finish LOD construction. 

Although \name takes 53s  to load the chunk points of the entire overview, users can already start exploring the data set as soon as the tile metadata is loaded. Chunkpoints are loaded progressively so users may navigate to already prepared regions, and a list of files/tiles allows users to zoom towards specific tiles, which are then loaded in full resolution even before all chunk points are loaded or heightmaps are generated.

\subsection{Surface Reconstruction}

\begin{table}[]
\caption{Root Mean Square Error (lower is better) of predicted heightmaps. Best results are in bold, second best underlined.}
\small
\setlength\tabcolsep{3.0pt}
\begin{tabular}{@{}lllll@{}}
\toprule
Data set   & Cubic              & HQSplat         & Linear             & \name                \\ \midrule
\dataCa    & 4.36±0.72          & 5.47±0.69       & {\ul 4.17±0.62}    & \textbf{3.81±0.55} \\
\dataSwiss & 11.60±2.14         & 12.58±2.16      & {\ul 9.26±1.44}    & \textbf{8.56±1.39} \\
\dataBora  & 1.40±0.82          & 1.97±0.31       & \textbf{1.00±0.60} & {\ul 1.16±0.17}    \\
\dataId    & \textbf{0.65±0.19} & 1.73±0.26       & {\ul 0.89±0.23}    & 1.22±0.21          \\
\dataNzA   & 11.38±4.71         & 7.43±0.93       & {\ul 6.97±1.12}    & \textbf{6.55±0.75} \\
\dataNzB   & 11.32±1.77         & {\ul 6.83±0.41} & 7.16±0.35          & \textbf{6.26±0.34} \\
\dataNzC   & 7.41±1.95          & 6.40±0.48       & {\ul 5.51±0.46}    & \textbf{5.09±0.33} \\ \midrule
Mean       & 6.87±1.76          & 6.06±0.75       & {\ul 4.99±0.69}    & \textbf{4.66±0.53} \\ \bottomrule
\end{tabular}
\label{tab:quant_comparison_hm}
\end{table}

\begin{table}[]
\caption{Peak Signal-To-Noise Ratio (higher is better) of predicted textures. Best results are in bold, second best underlined.}
\small
\setlength\tabcolsep{3.0pt}
\begin{tabular}{@{}lllll@{}}
\toprule
Data set   & Cubic      & HQSplat             & Linear     & \name                 \\ \midrule
\dataCa    & 66.08±1.26 & {\ul 69.83±1.14}    & 69.38±1.14 & \textbf{70.76±1.09} \\
\dataBora  & 69.57±1.11 & \textbf{78.42±1.19} & 73.29±0.97 & {\ul 77.45±0.86}    \\
\dataId    & 67.60±3.44 & \textbf{78.79±0.80} & 75.02±0.53 & {\ul 77.89±0.87}    \\
\dataNzA   & 66.98±1.15 & {\ul 71.30±1.03}    & 70.80±0.95 & \textbf{72.24±0.77} \\
\dataNzB   & 64.54±0.52 & {\ul 70.86±0.63}    & 68.67±0.54 & \textbf{71.46±0.48} \\
\dataNzC   & 65.84±0.67 & {\ul 70.47±0.68}    & 69.67±0.57 & \textbf{71.22±0.52} \\ \midrule
Mean       & 66.77±1.36 & {\ul 73.28±0.91}    & 71.14±0.79 & \textbf{73.50±0.77} \\ \bottomrule
\end{tabular}
\label{tab:quant_comparison_rgb}
\end{table}

\begin{table*}[]
\caption{Ablation study main results. Best results are in bold, second best underlined. Note that \diff{the \textbf{Extra Data} variant was partially trained on the test data, so its metrics are positively biased.} Please see the supplementary material for the full tables.}
\setlength\tabcolsep{3.0pt}
\centering
\begin{tabular}{@{}l|llllllll@{}}
\toprule
Variant            & NPBG       & DCTNet     & Raster     & HM Only         & NN only    & Lin only         & \diff{Extra Data}             & \name      \\ \midrule
Heights RMSE [m] ↓ & 4.87±0.64  & 4.82±0.56  & 14.91±3.30 & {\ul 4.64±0.55} & 4.84±0.55  & 4.71±0.55        & \textbf{4.56±0.52}  & 4.66±0.53  \\
Colors PSNR [DB] ↑ & 73.30±0.77 & 73.20±0.77 & 60.98±1.07 & NA              & 73.50±0.77 & {\ul 73.53±0.76} & \textbf{74.16±0.78} & 73.50±0.77 \\ \bottomrule
\end{tabular}
\label{tab:ablation}
\end{table*}



\begin{figure*}[h]
\centering
\includegraphics[width=\linewidth]{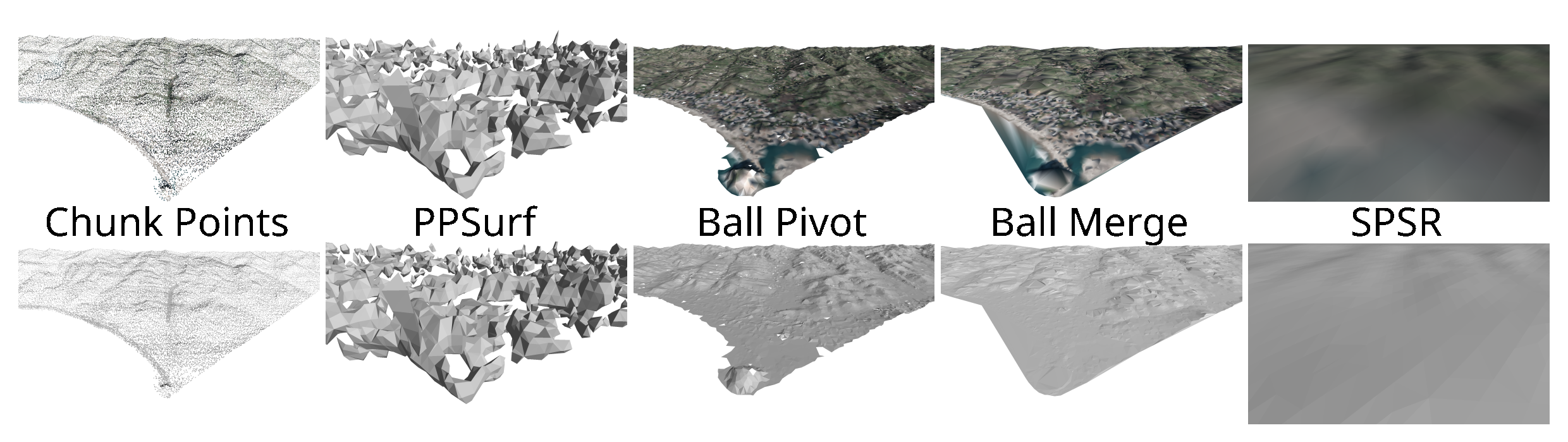}
\caption{\diff{Qualitative comparison of the Morro Rock region in \dataCa.}}
\label{fig:qual_comp}
\end{figure*}

\diff{Few methods for surface reconstruction are applicable to our use case. Recent and popular global reconstruction methods, such as BallMerge~\cite{parakkat2024ballmerge}, Ball Pivot~\cite{bernardini1999ball}, Screened Poisson Surface Reconstruction~\cite{kazhdan2013screened}, and PPSurf~\cite{erler2024ppsurf} perform poorly with our chunk points. Please see the supplementary material for images. Figure~\ref{fig:qual_comp} shows a qualitative comparison of the most relevant local reconstruction methods. Linear interpolation has hard discontinuities and noisy colors (see also Fig.~\ref{fig:patch_example}). Cubic interpolation suffers from overshooting, causing very bright and dark spots, and sometimes peaks of several hundred meters. High-Quality Splatting~\cite{botsch2005high} either creates blobby structures, smoothed stairs, or gaps due to the fixed-size kernel. NPBG~\cite{aliev2020neural} with our inputs is close to \name in quality but distorts colors sometimes.}



\paragraph*{Quantitative Comparison}



Tables~\ref{tab:quant_comparison_hm} and~\ref{tab:quant_comparison_rgb} show the performance of \name on all data sets. We report the average over all patches in the test sets. We use Root Mean Squared Error (RMSE) in meters to compare heightmap quality and Peak Signal-to-Noise Ratio (PSNR) in dB for textures.
 
Our most important baseline is also used as input to our network: linearly interpolated heightmaps and textures. We perform Delaunay triangulation and linear interpolation with barycentric coordinates on the local chunk points, as described in Section~\ref{sec:method_raw_hm}. For such a simple baseline, it is surprisingly good. It is also an approximation for Rapidlasso's Las2Dem\cite{rapidlassoGeneratingSpikeFree}, which takes care of multiple returns per texels in addition. However, this refinement does not make a noticeable difference with our sparse chunk points. We also compare with a cubic Clough-Tocher interpolation implemented in SciPy~\cite{scipy2024ct2dinterp}, which is accurate in many cases but occasionally overshoots. Lastly, we compare to High-Quality Splatting~\cite{botsch2005high} by treating each chunk point as a large, fixed-size splat, and using a Gaussian blending function to obtain a smooth transition of heights and colors between overlapping splats. 
\name performs significantly better than the baselines except for the much denser \dataBora and \dataId data sets. 




\paragraph*{Computation Time and Memory Consumption}



Reconstruction was evaluated on an NVIDIA RTX 3090 and an AMD Ryzen 7 3700X 8-Core. The reconstruction of one patch in our C++/CUDA framework takes around 20 ms. Single-threaded triangulation and sampling take 16 ms, inference and buffer copies 4 ms. Batched inference requires copying data to make the input heightmaps and textures contiguous in memory, which means a small overhead. In any case, the timings (see supplementary material) show that batching always pays off.

\paragraph*{Ablation}


Table~\ref{tab:ablation} shows an ablation that empirically validates our design choices, comparing different network architectures and inputs.  Other architectures like \textbf{NPBG}~\cite{aliev2020neural} and \textbf{DCTNet}~\cite{zhao2022discrete} perform worse than ours. \textbf{Raster}izing points and filling unknown pixels with zeros does not work well as input for our image-based network. This means that dense inputs are necessary for viable quality.
The model draws information from both nearest-neighbor (\textbf{NN}) and \textbf{lin}ear interpolation inputs, especially for the heights. \textbf{Lin}ear-only generalizes better across point densities, while \textbf{NN}-only is better with similar densities. Combining them is a step towards the best of both. Omitting RGB inputs (\textbf{HM only}) has a negligible impact on heightmap quality. \diff{Adding \textbf{Extra Data} (\dataId and \dataNz in addition to \dataCa and \dataSwiss) improves the quality significantly. Note that only \dataBora is completely unseen, making a fair comparison difficult for this variant.} Please see the supplementary material for detailed statistics.

\subsection{Discussion and Limitations}

\diff{Lidarserv~\shortcite{RealTimeIndexingBormann} and SimLOD~\cite{schutz2024simlod} are prior work that follow similar goals: Exploring large data sets without the need to preprocess and wait. Lidarserv specifically aims to enable displaying arbitrarily large point clouds directly during capture, and is able to construct out-of-core LOD structures at rates of up to around 1.8 million points per second. SimLOD aims to visualize large point cloud files quickly and is capable of loading industry standard LAS files at rates of up to 300 million points per second, or compressed LAZ files at up to 30 million points per second. Both display points immediately as they are streamed, without the need to wait until processing is finished. A major difference in our approach is that we aim to display massive data sets in their entirety in a matter of seconds and prioritize more detailed reconstructions towards the user's viewpoint, while the prior works operate on local regions in undefined order without priorization. SimLOD is further limited to data sets that fit in memory, i.e., about 800 million points per 24GB of memory. Our approach, on the other hand, rapidly displays arbitrarily large data sets but lacks level-of-detail structures that would further improve rendering performance, especially for previously visited regions. In the future, we would like to integrate and expand SimLOD's incremental LOD construction in order to build an out-of-core system that is capable of rendering arbitrarily large point clouds with instant overviews of the entire data set, and priorization towards the current viewpoint. }

\begin{figure*}[ht]
\centering
\includegraphics[width=0.95\linewidth]{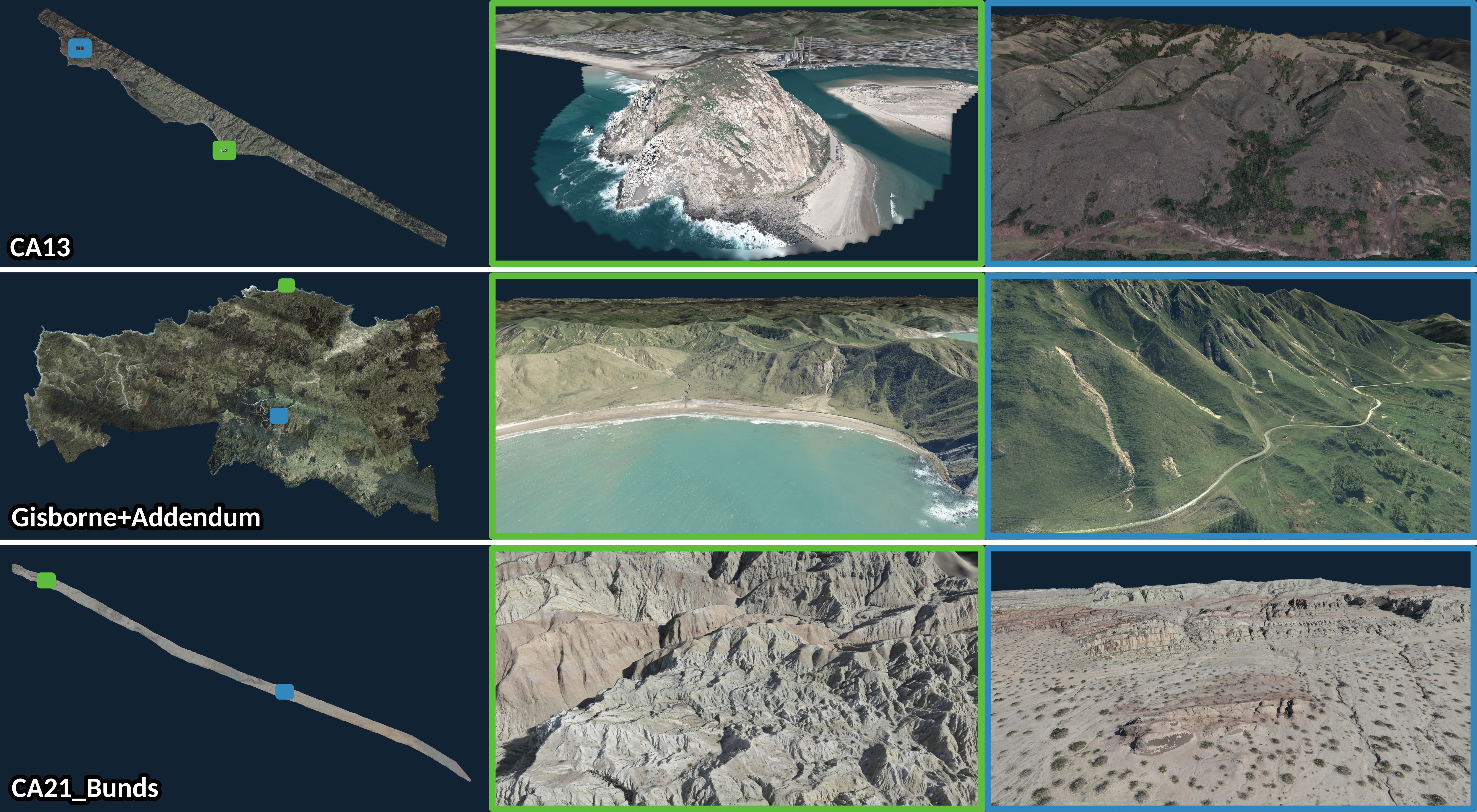}
\caption{Screenshots of CA13 (17.7B points, 90GB), Gisborne (262B points, 2.4TB) and CA21\_Bunds (8.4B points, 96GB) made with \name.}
\label{fig:screenshots}
\end{figure*}

Although we trained \name only on \dataCa and \dataSwiss, it generalizes well to other regions of the world. The model has mostly seen colors typical for deserts, small cities, and beaches from the Morro Bay region in California, USA. Nonetheless, it can still reconstruct the colors of the lush vegetation of Gisborne, New Zealand, as shown in Table~\ref{tab:quant_comparison_rgb} and Figure~\ref{fig:screenshots}. Furthermore, it generalizes well across scan patterns that may affect the sparse subsample. It was trained on the line-wise scanning patterns of the \dataCa aerial LIDAR, but it has no issues with the circular scanning patterns of \dataNz. 

Linear and cubic interpolation are competitive on the much denser photogrammetry point clouds of \dataId and \dataBora for predicting heights. However, only HQ-Splatting can compete with our method when estimating colors. This indicates that our method does not generalize too well from point densities of around 20 $points/m^2$ in \dataCa to almost 600 $points/m^2$ in \dataBora and fails to use the available information. Adapting the patch size solves this issue. 

\section{Conclusion}

\name is the first point cloud viewer that allows exploring terabytes of compressed LIDAR scans within seconds without any pre-processing. We present fast 2.5D surface reconstruction from sparse, local subsamples with minimal overhead. Our neural network outperforms the current industry standard of triangulation and linear interpolation. In the future, the local chunk points or heightmaps could be streamed from a server for only the required regions, which would drastically reduce data storage on the user side and data transfer costs for everyone.

Our approach is potentially applicable to any data that allows at least partial random access with some learnable patterns, such as huge photographs, volumetric data from CT or MRT scans, weather forecasts, and astronomy simulations. The neural network can be adapted for annotation, segmentation, and classification tasks. With the latter, for example, it could take the number of returns and other extra point properties to detect vegetation. Extending \name to 3D for large urban and indoor scans should be possible by combining our persistent heightmaps with screen-space approaches like ADOP~\cite{ruckert2022adop} and TRIPS~\cite{franke2024trips}.



\section{Acknowledgements}

The authors wish to thank following data set providers: \emph{Bunds at el.} and \emph{Open Topography} for the Bund\_Bora~\shortcite{Bund_BoraPk} and ID15\_Bunds~\shortcite{ID15_Bunds} data sets; 
\emph{PG\&E} and \emph{Open Topography} for CA13~\shortcite{CA13_SAN_SIM}; 
The \emph{Ministry of Business, Innovation and Employment} and \emph{Toitū Te Whenua Land Information New Zealand} and \emph{Open Topography} for Gisborne~\shortcite{NZ23_Gisborne}; 
The \emph{São Paulo City Hall (PMSP)} and \emph{Open Topography} for São Paulo~\shortcite{sao_paulo};
The \emph{Bundesamt für Landestopografie swisstopo} for swissSURFACE3D~\cite{swissSURFACE3Draster}.


We thank Paul Guerrero, Pedro Hermosilla, and Adam Celarek for their valuable inputs. Further, we thank Stefan Ohrhallinger for running reconstructions with BallMerge~\cite{parakkat2024ballmerge}.

This research has been funded by WWTF project \emph{ICT22-055 - Instant Visualization and Interaction for Large Point Clouds}. 



\begin{figure}[t]
    \centering
    \begin{subfigure}[b]{0.45\textwidth}
        \centering
        \includegraphics[width=\textwidth]{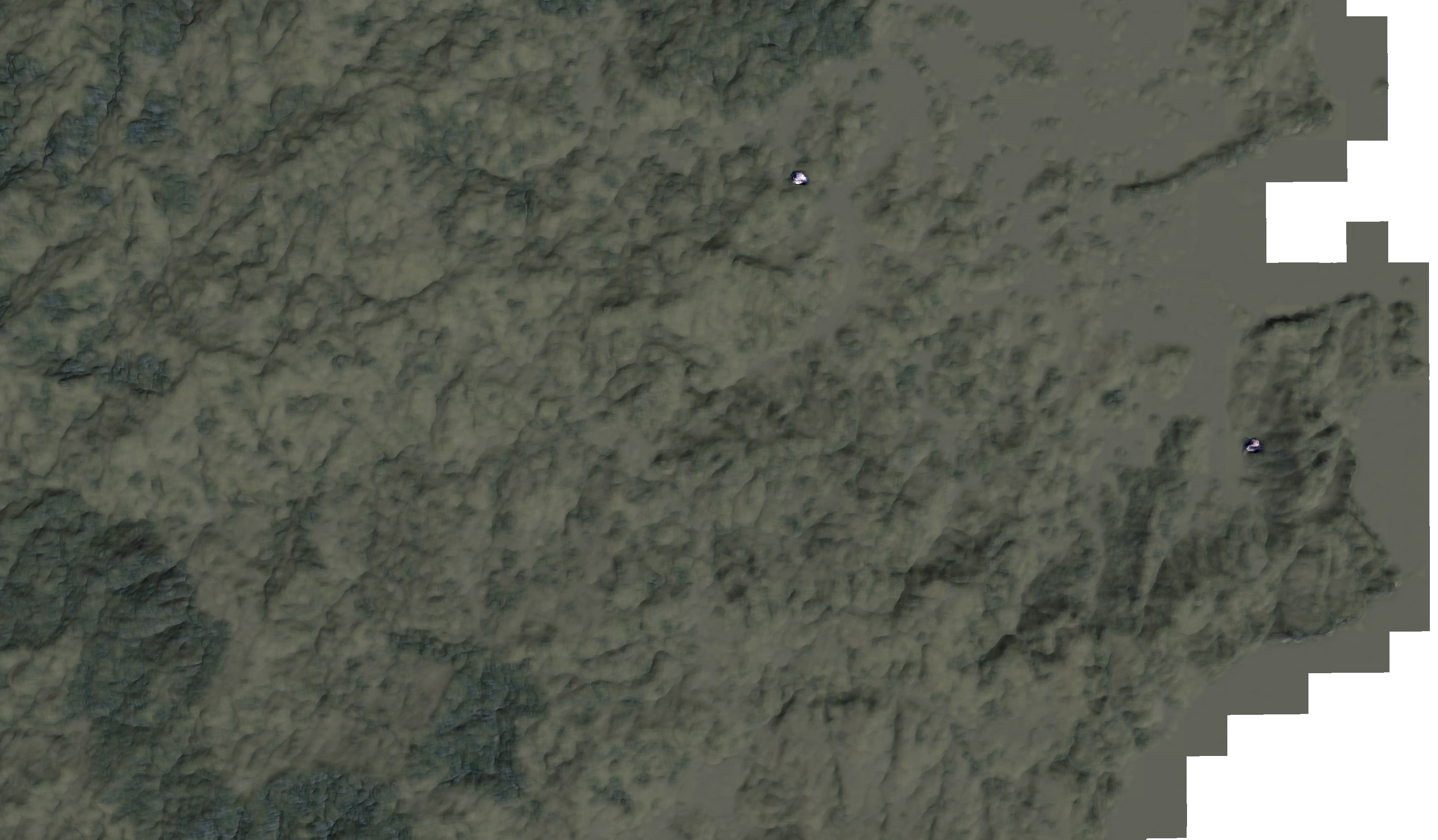}
        \caption{Colorizer}
        \label{fig:sub1}
    \end{subfigure}
    \hfill
    \begin{subfigure}[b]{0.45\textwidth}
        \centering
        \includegraphics[width=\textwidth]{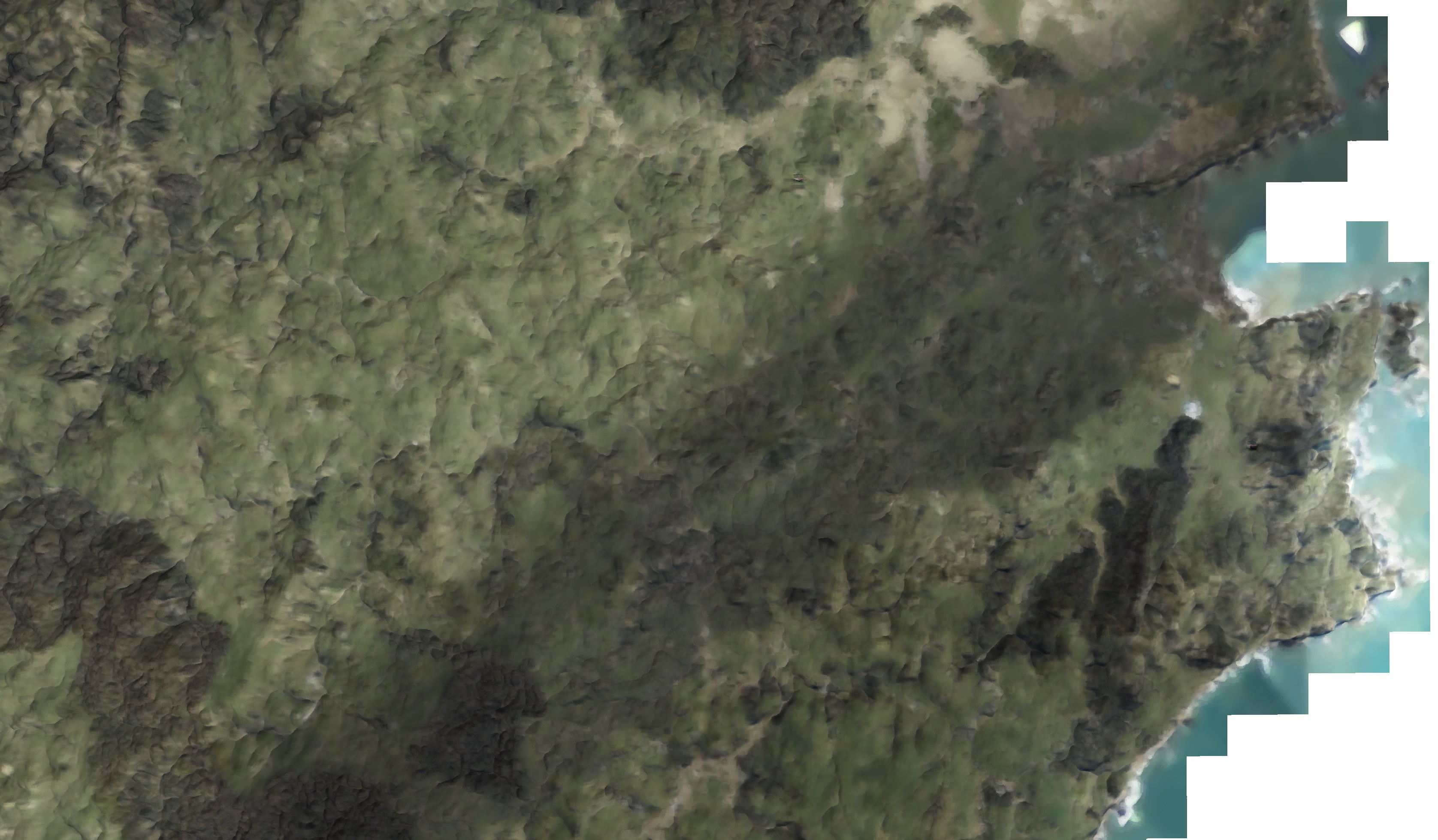}
        \caption{\diff{\name Extra Data}}
        \label{fig:sub2}
    \end{subfigure}
    \caption{Colorization result. The colorizer model did not receive RGB inputs but was forced to output colors. }
    \label{fig:colorization}
\end{figure}


\begin{figure}[t]
    \centering
    \includegraphics[width=\linewidth]{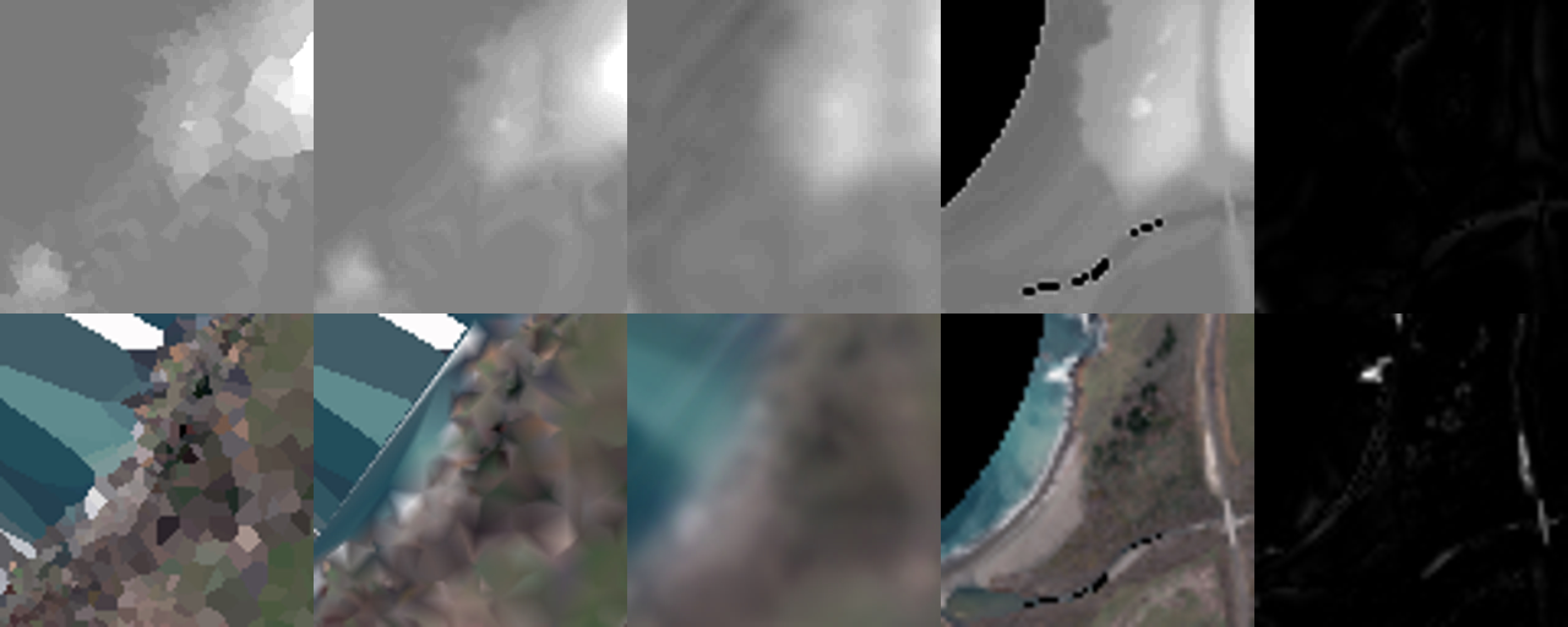}
    \caption{Example patch. The top row contains heightmaps and the bottom row textures. Left to right: Nearest neighbor (96x96), linear (96x96), network output (64x64), ground-truth (64x64), loss (64x64). Black regions in ground-truth are missing areas in the scan. Note the high RGB loss in the shallow water near the road where only a single chunk point contained relevant information. Scan boundaries usually come with slender triangles and Voronoi-like regions. }
    \label{fig:patch_inputs_pred_los}
\end{figure}


\begin{table}[t]
\caption{Comparison of inference times in C++/CUDA with different batch sizes on an NVIDIA RTX 3090 with an AMD Ryzen 7 3700X. The numbers shown are median over 1000 runs in milliseconds of inference calls surrounded by device synchronization. Batch size None is the unbatched version of the inference with a bit less overhead per call but scales linearly with the number of processed tiles. Batching always pays off.}
\centering 
\footnotesize
\setlength{\tabcolsep}{2.5pt}
\begin{tabular}{@{}lll@{}}
\toprule
Batch Size & Time {[}ms{]}           & Per Tile {[}ms{]}      \\ \midrule
None       & 4.2                     & 4.2                    \\ \midrule
1          & 4.2                     & 4.2                    \\
2          & 4.3                     & 2.2                    \\
3          & 4.6                     & 1.5                    \\
4          & 4.7                     & 1.2                    \\
5          & 4.9                     & 1.0                    \\
10         & 6.4                     & 0.6                    \\ \bottomrule
\end{tabular}
\end{table}
\label{tab:timings_batched}


\begin{table*}[t]
\caption{Ablation study. We show the RMSE of the heightmaps produced by \name with various changes. Note that Linear interpolation is still better than Extra Data for reconstructing heights (not colors) of the photogrammetry datasets. This indicates a significant bias towards typical point densities of LIDAR scans, which \name cannot fully generalize across. For future work, we recommend using separate models for LIDAR and photogrammetry, adapting the meters per pixel, or training with varying sampling densities. }
\begin{tabular}{@{}lllllllll@{}}
\toprule
Dataset    & NPBG      & DCTNet             & HM Only            & Rast        & NN only         & Lin only  & Extra Data      & \name           \\ \midrule
\dataCa    & 3.92±0.53 & \textbf{3.80±0.55} & 3.84±0.55          & 6.04±0.85   & 4.10±0.55       & 3.84±0.56 & 3.82±0.54          & {\ul 3.81±0.55} \\
\dataSwiss & 9.25±2.01 & 8.87±1.52          & 8.70±1.43          & 14.07±2.96  & 9.34±1.55       & 8.58±1.38 & \textbf{8.55±1.39} & {\ul 8.56±1.39} \\
\dataBora  & 1.39±0.19 & 1.55±0.21          & \textbf{1.13±0.19} & 39.32±11.50 & 1.28±0.20       & 1.27±0.18 & \textbf{1.13±0.17} & {\ul 1.16±0.17} \\
\dataId    & 1.37±0.18 & 1.54±0.31          & {\ul 1.19±0.27}    & 21.88±5.62  & 1.21±0.17       & 1.29±0.28 & \textbf{1.05±0.18} & 1.22±0.21       \\
\dataNzA   & 6.70±0.87 & 6.48±0.66          & {\ul 6.44±0.72}    & 8.39±1.16   & 6.55±0.68       & 6.60±0.76 & \textbf{6.40±0.72} & 6.55±0.75       \\
\dataNzB   & 6.31±0.35 & 6.32±0.34          & 6.19±0.35          & 7.65±0.34   & {\ul 6.14±0.35} & 6.26±0.34 & \textbf{6.00±0.34} & 6.26±0.34       \\
\dataNzC   & 5.14±0.33 & 5.18±0.33          & 5.03±0.34          & 7.05±0.67   & 5.25±0.33       & 5.12±0.34 & \textbf{4.98±0.32} & {\ul 5.09±0.33} \\ \midrule
Mean       & 4.87±0.64 & 4.82±0.56          & {\ul4.64±0.55}     & 14.91±3.30  & 4.84±0.55       & 4.71±0.55 & \textbf{4.56±0.52} &  4.66±0.53       \\ \bottomrule
\end{tabular}
\end{table*}

\begin{table*}[t]
\caption{Ablation study. We show the PSNR of the textures produced by \name with various changes.}
\begin{tabular}{@{}lllllllll@{}}
\toprule
Dataset    & NPBG       & DCTNet           & HM Only & Rast       & NN only          & Lin only         & Extra Data       & \name               \\ \midrule
\dataCa    & 70.61±1.05 & 70.57±1.05       & NA      & 69.61±0.99 & 70.68±1.08       & 70.56±1.06       & {\ul 70.74±1.10}    & \textbf{70.76±1.09} \\
\dataSwiss & NA         & NA               & NA      & NA         & NA               & NA               & NA                  & NA                  \\
\dataBora  & 77.03±0.83 & 76.57±0.88       & NA      & 40.87±1.57 & 77.31±0.89       & {\ul 77.67±0.88} & \textbf{78.20±0.91} & 77.45±0.86          \\
\dataId    & 77.48±0.82 & 77.07±0.89       & NA      & 45.08±1.74 & 77.72±0.81       & {\ul 78.13±0.82} & \textbf{79.26±0.76} & 77.89±0.87          \\
\dataNzA   & 72.07±0.91 & 72.20±0.78       & NA      & 70.14±1.15 & {\ul 72.31±0.78} & 72.24±0.77       & \textbf{72.57±0.77} & 72.24±0.77          \\
\dataNzB   & 71.58±0.49 & 71.51±0.47       & NA      & 70.29±0.53 & {\ul 71.69±0.50} & 71.45±0.49       & \textbf{72.42±0.55} & 71.46±0.48          \\
\dataNzC   & 71.04±0.50 & {\ul 71.27±0.53} & NA      & 69.87±0.45 & {\ul 71.27±0.53} & 71.17±0.53       & \textbf{71.77±0.57} & 71.22±0.52          \\ \midrule
Mean       & 73.30±0.77 & 73.20±0.77       & NA      & 60.98±1.07 & 73.50±0.77       & {\ul 73.53±0.76} & \textbf{74.16±0.78} & 73.50±0.77          \\ \bottomrule
\end{tabular}
\end{table*}


\begin{table*}[t]  
\small
\caption{Colorization quality. The colorizer model was trained on the same datasets as the Extra Data ablation model but did not receive color inputs. As expected, the results of the colorizer are significantly worse, even the heightmaps are a bit worse. In the future, a generative model running on larger regions could produce good results. }
\begin{tabular}{@{}lllll@{}}
\toprule
           & \multicolumn{2}{c}{Heights RMSE [m] ↓}  & \multicolumn{2}{c}{Colors PSNR [DB] ↑} \\ \midrule
Dataset    & Colorizer          & Extra Data         & Colorizer     & Extra Data            \\
\dataCa    & 3.83±0.55          & \textbf{3.82±0.54} & 67.04±0.93    & \textbf{70.74±1.10}    \\
\dataSwiss & 8.69±1.39          & \textbf{8.55±1.39} & NA            & NA                     \\
\dataBora  & \textbf{1.13±0.20} & \textbf{1.13±0.17} & 69.32±1.29    & \textbf{78.20±0.91}    \\
\dataId    & 1.23±0.25          & \textbf{1.05±0.18} & 69.23±1.20    & \textbf{79.26±0.76}    \\
\dataNzA   & 6.45±0.74          & \textbf{6.40±0.72} & 66.96±1.43    & \textbf{72.57±0.77}    \\
\dataNzB   & 6.23±0.34          & \textbf{6.00±0.34} & 68.40±0.78    & \textbf{72.42±0.55}    \\
\dataNzC   & 5.03±0.33          & \textbf{4.98±0.32} & 67.22±0.68    & \textbf{71.77±0.57}    \\ \midrule
Mean       & 4.66±0.54          & \textbf{4.56±0.52} & 68.03±1.05    & \textbf{74.16±0.78}    \\ \bottomrule
\end{tabular}
\end{table*}



\begin{figure*}[t]
    \centering
    \begin{subfigure}[b]{0.23\textwidth}
        \centering
        \includegraphics[width=\textwidth]{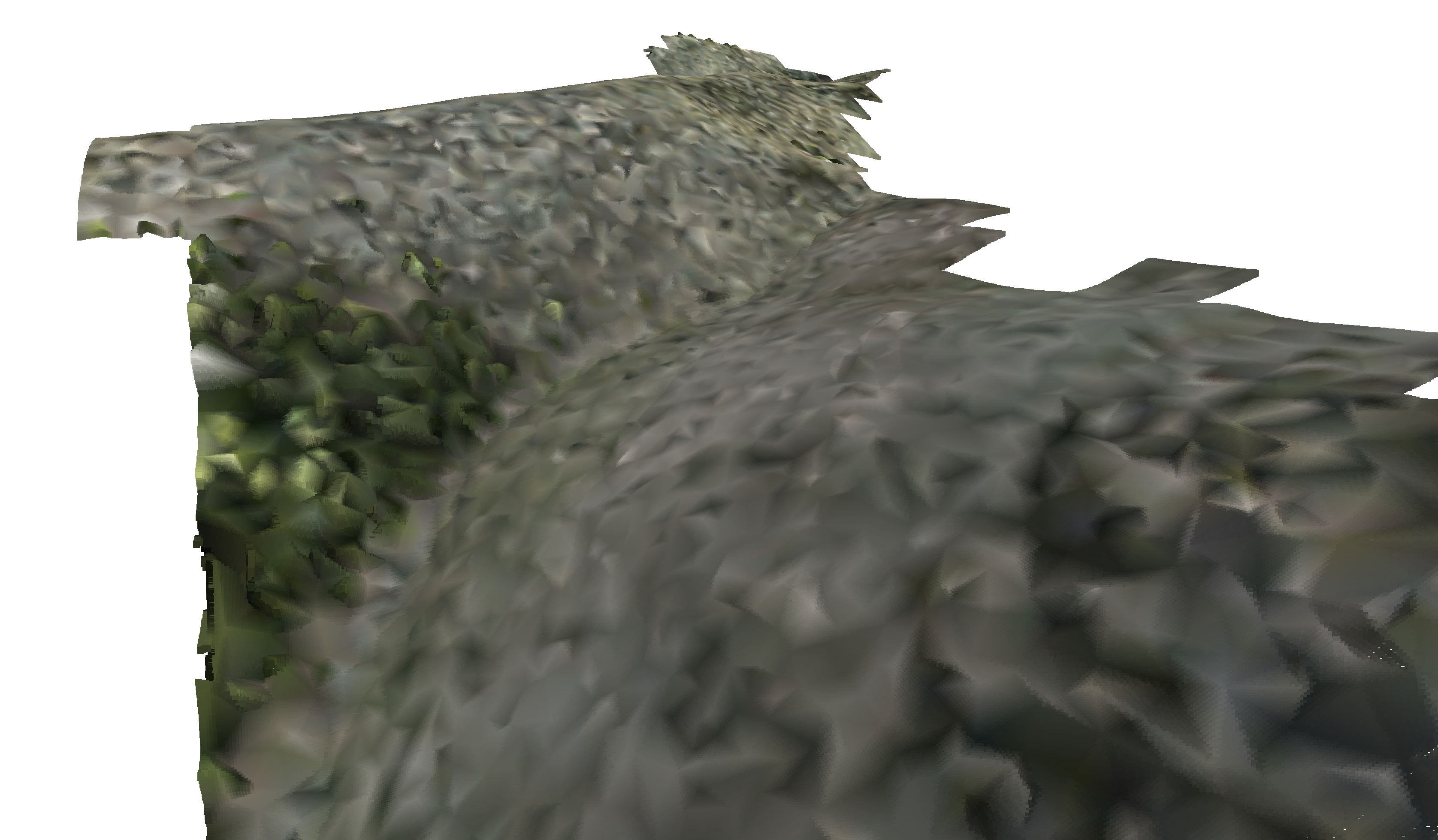}
    \end{subfigure}
    \hfill
    \begin{subfigure}[b]{0.23\textwidth}
        \centering
        \includegraphics[width=\textwidth]{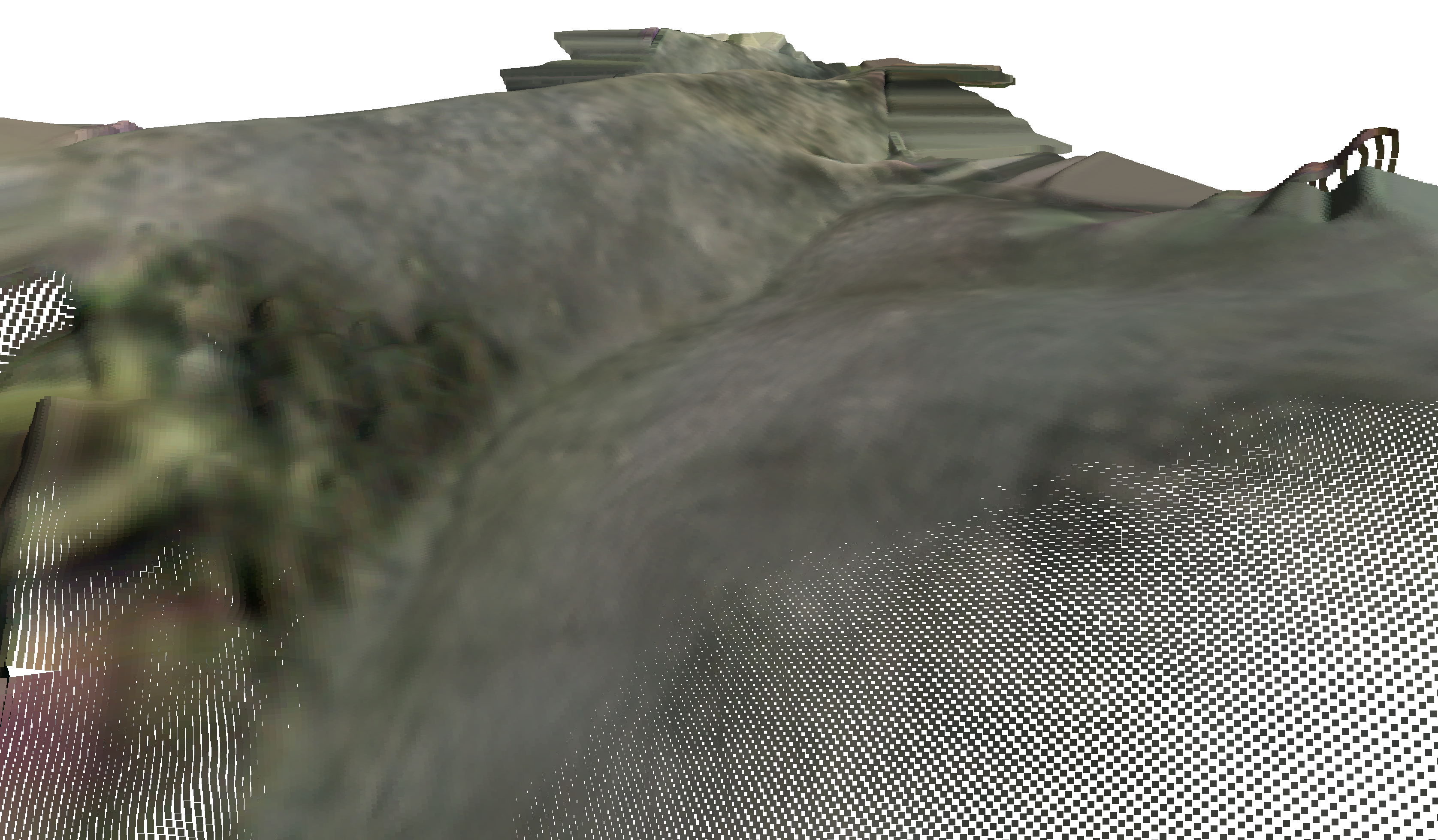}
    \end{subfigure}
    \hfill
    \begin{subfigure}[b]{0.23\textwidth}
        \centering
        \includegraphics[width=\textwidth]{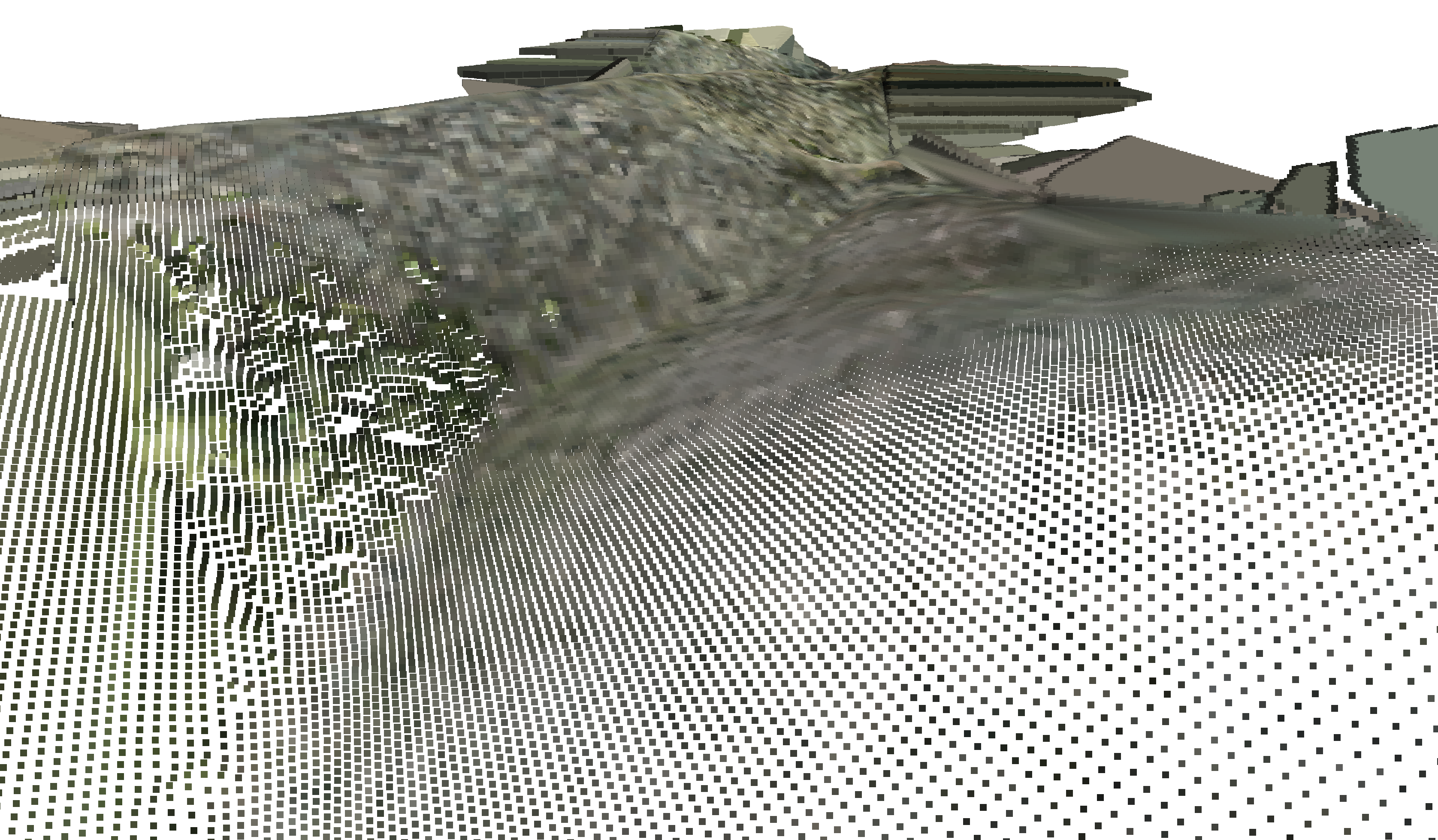}
    \end{subfigure}
    \hfill
    \begin{subfigure}[b]{0.23\textwidth}
        \centering
        \includegraphics[width=\textwidth]{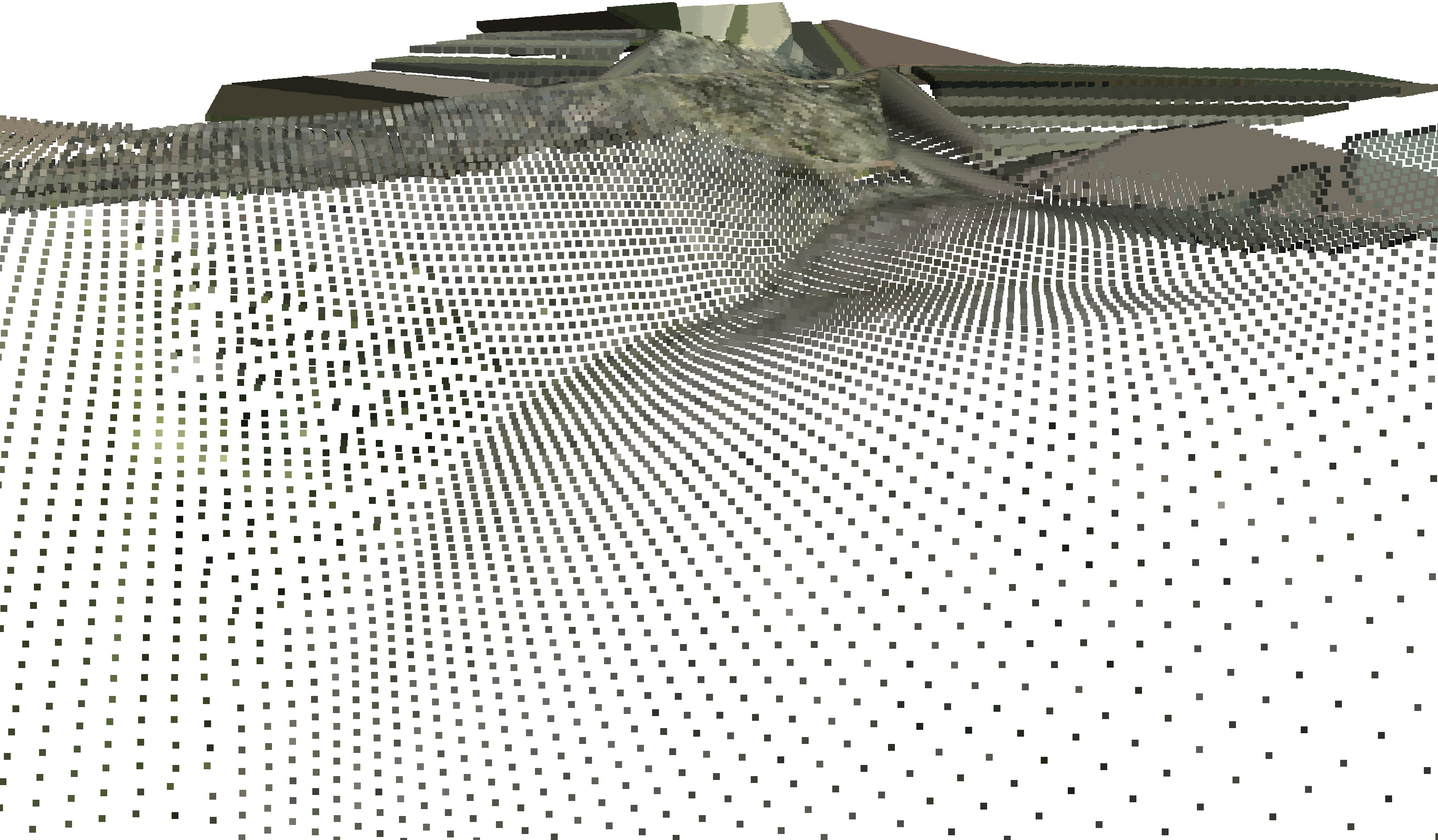}
    \end{subfigure}

    \vspace{1em} 

    \begin{subfigure}[b]{0.23\textwidth}
        \centering
        \includegraphics[width=\textwidth]{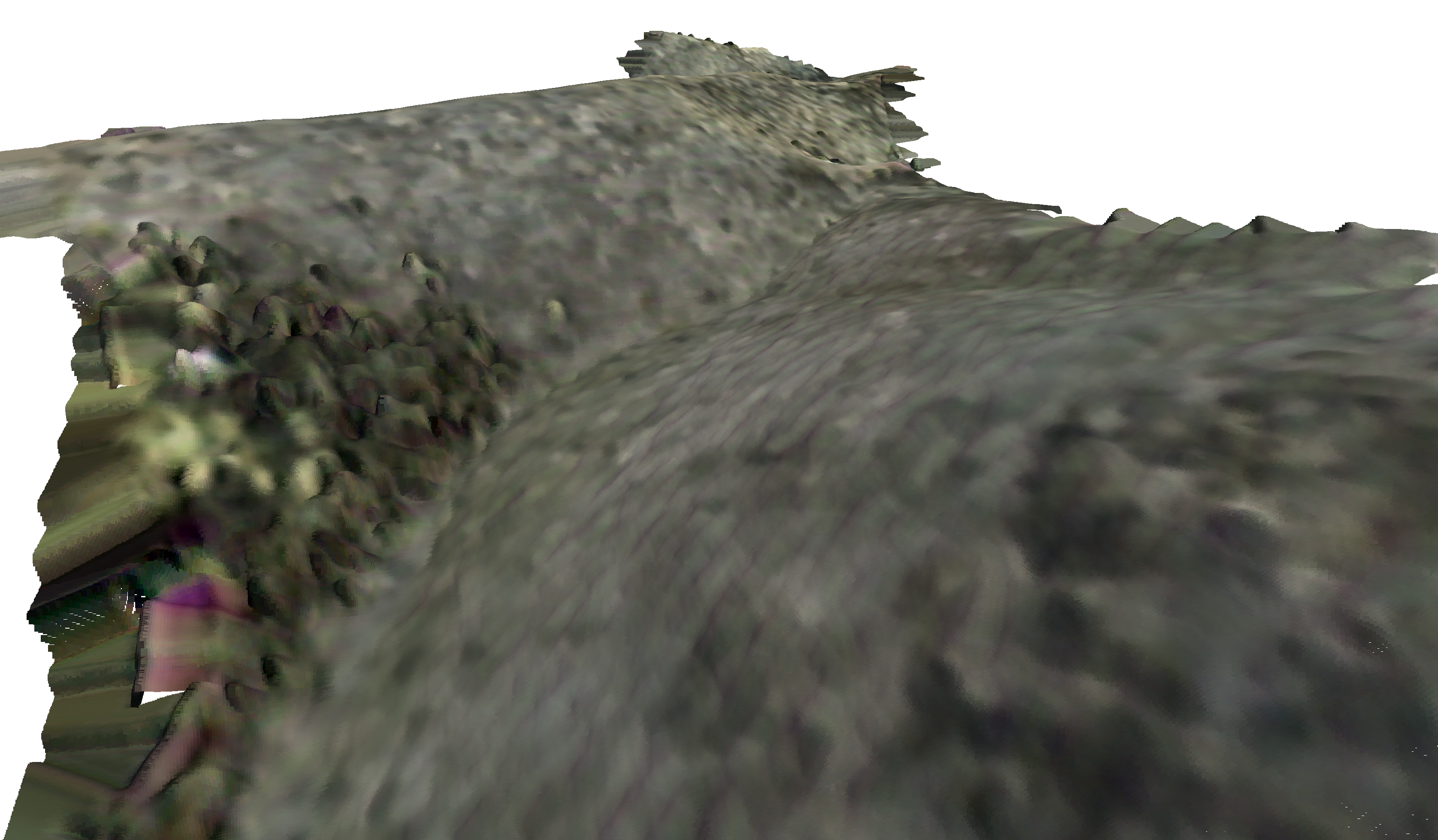}
    \end{subfigure}
    \hfill
    \begin{subfigure}[b]{0.23\textwidth}
        \centering
        \includegraphics[width=\textwidth]{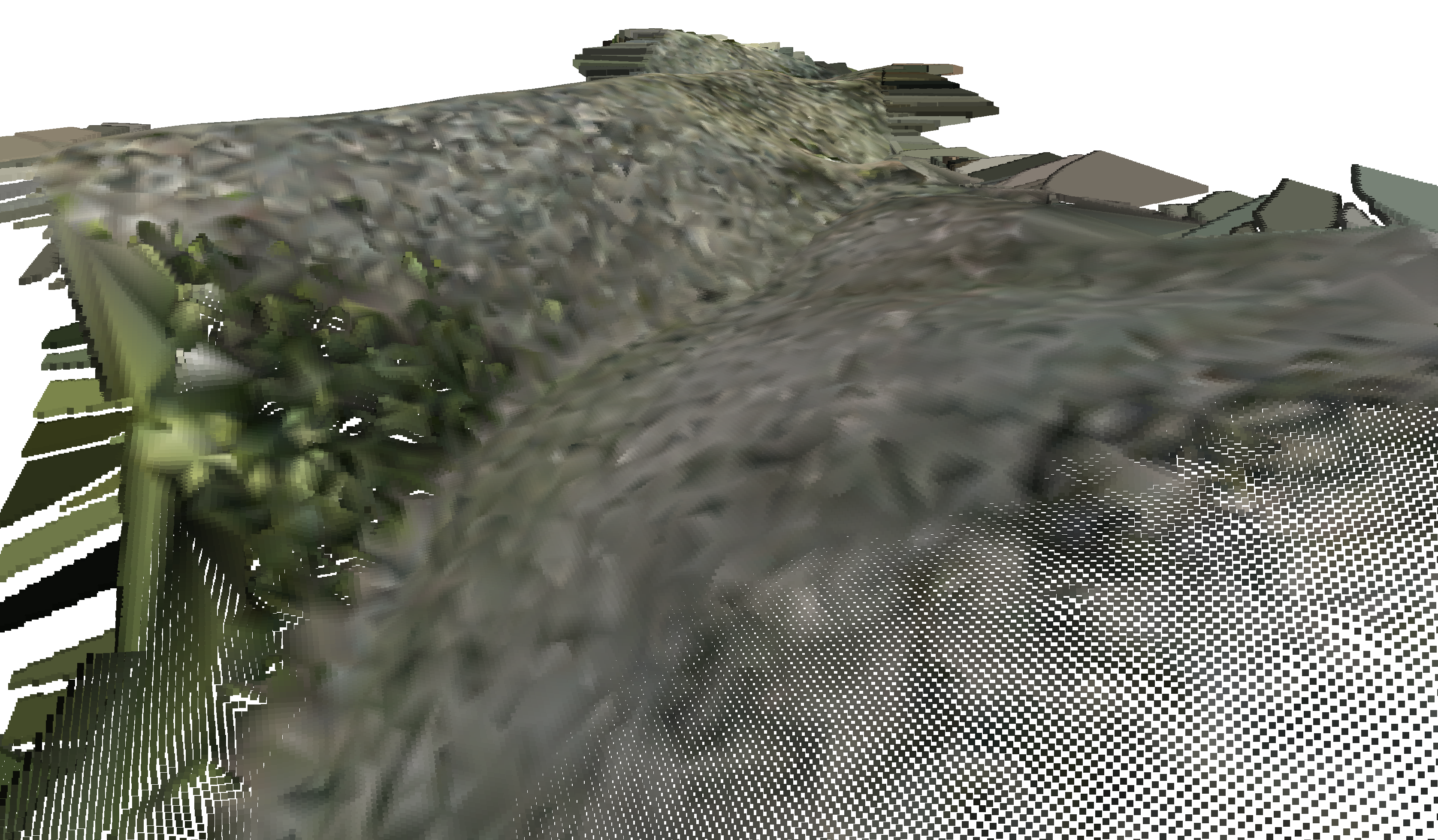}
    \end{subfigure}
    \hfill
    \begin{subfigure}[b]{0.23\textwidth}
        \centering
        \includegraphics[width=\textwidth]{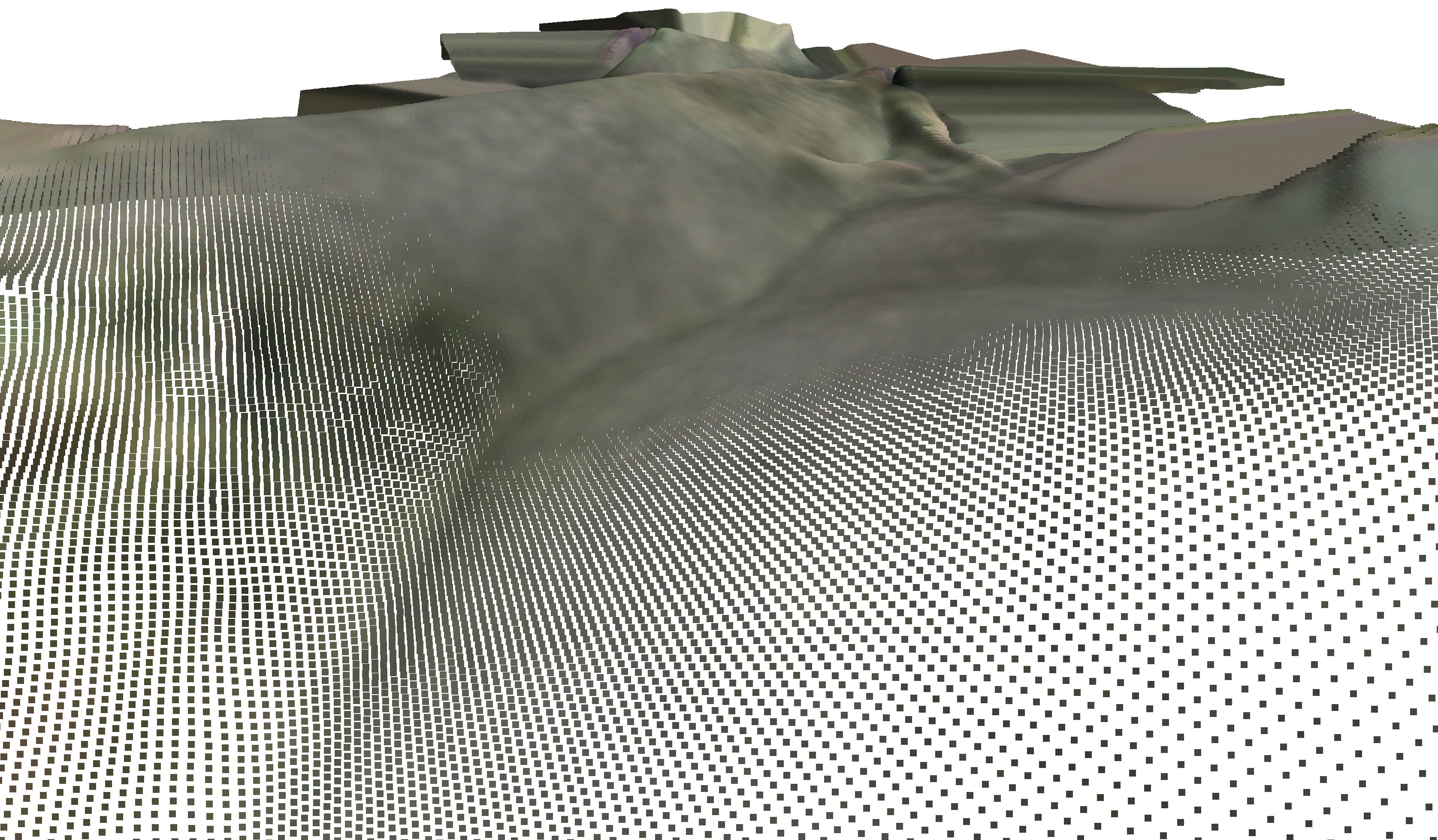}
    \end{subfigure}
    \hfill
    \begin{subfigure}[b]{0.23\textwidth}
        \centering
        \includegraphics[width=\textwidth]{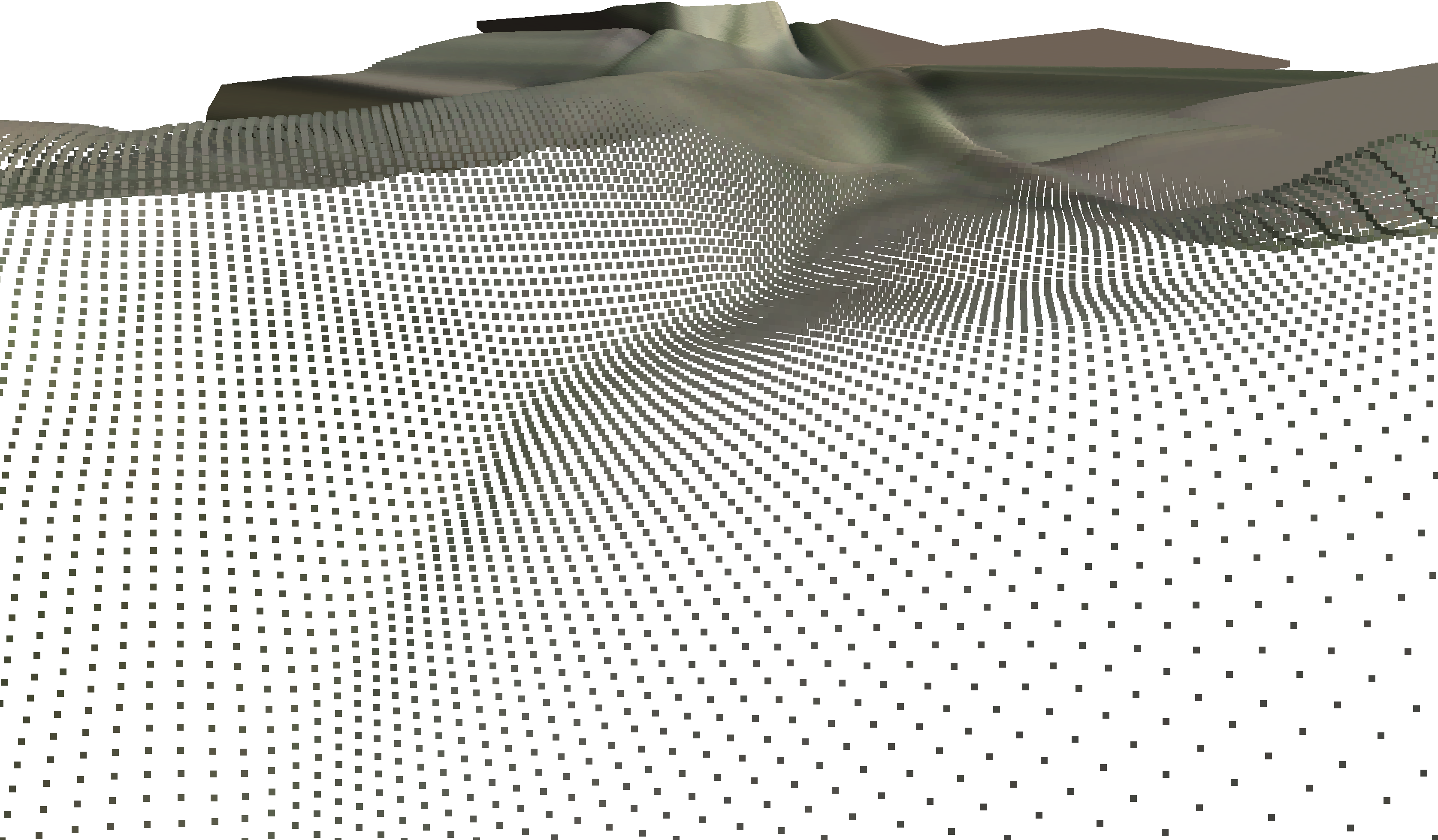}
    \end{subfigure}
    
    \caption{Reconstruction of \dataBora with different resolutions. Upper: linear interpolation. Lower: \name. Left to right: 1, 2.5, 5, 10 (default) meters per texel. }
    \label{fig:hm_res_bora}
\end{figure*}


\begin{figure*}[t]
    \centering
    \begin{subfigure}[b]{0.30\textwidth}
        \centering
        \includegraphics[width=\textwidth]{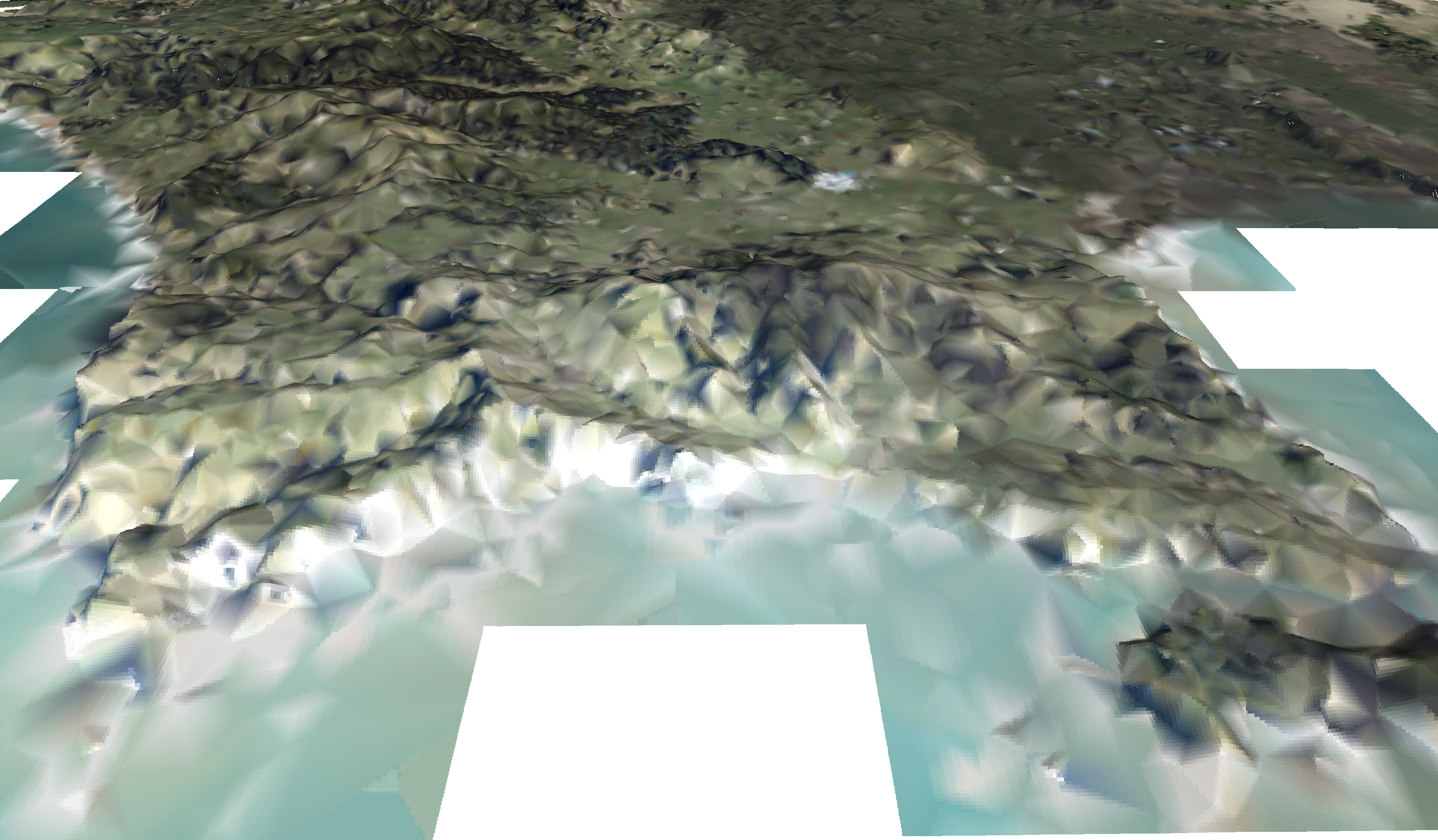}
    \end{subfigure}
    \hfill
    \begin{subfigure}[b]{0.30\textwidth}
        \centering
        \includegraphics[width=\textwidth]{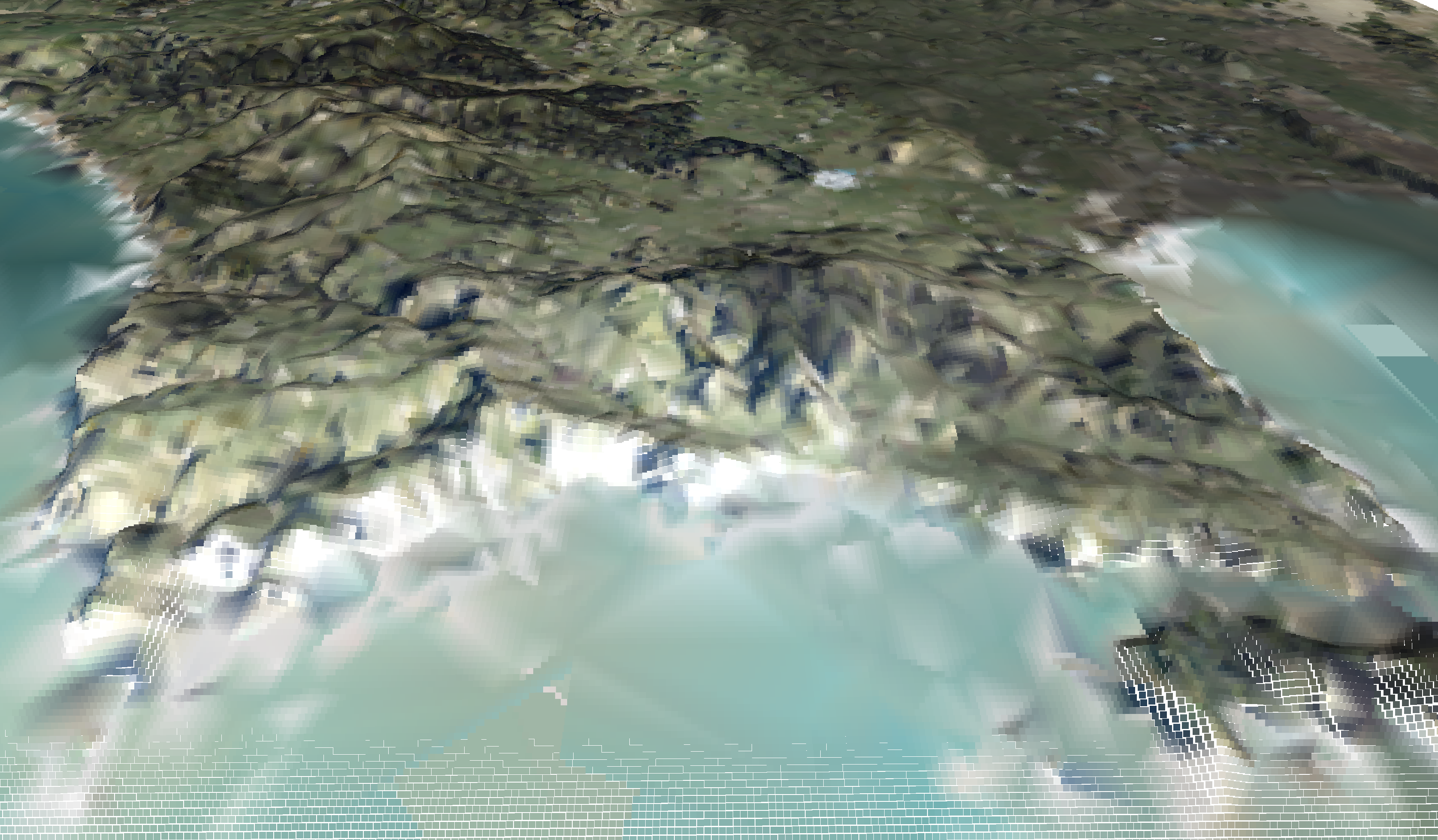}
    \end{subfigure}
    \hfill
    \begin{subfigure}[b]{0.30\textwidth}
        \centering
        \includegraphics[width=\textwidth]{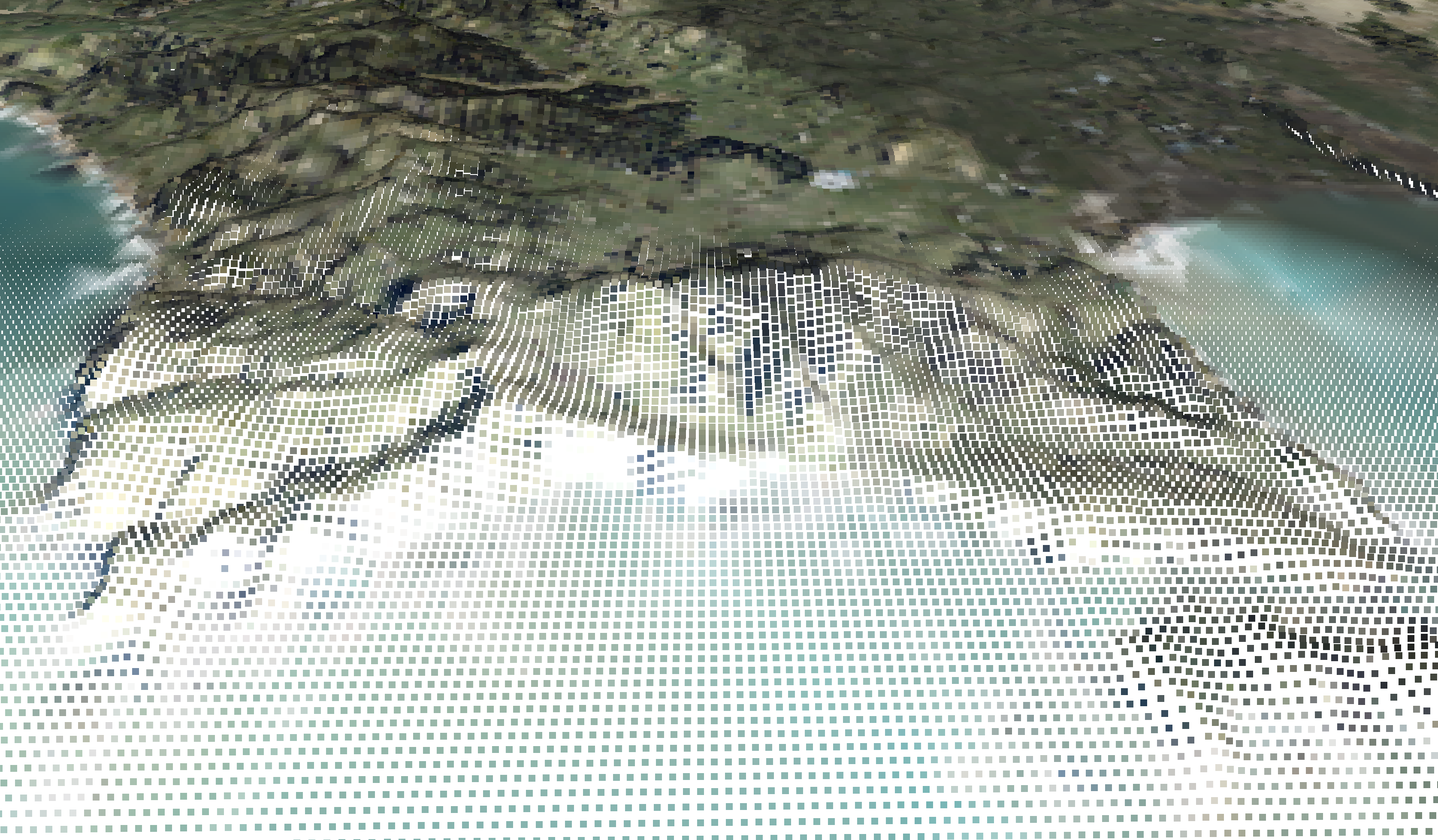}
    \end{subfigure}

    \vspace{1em} 

    \begin{subfigure}[b]{0.30\textwidth}
        \centering
        \includegraphics[width=\textwidth]{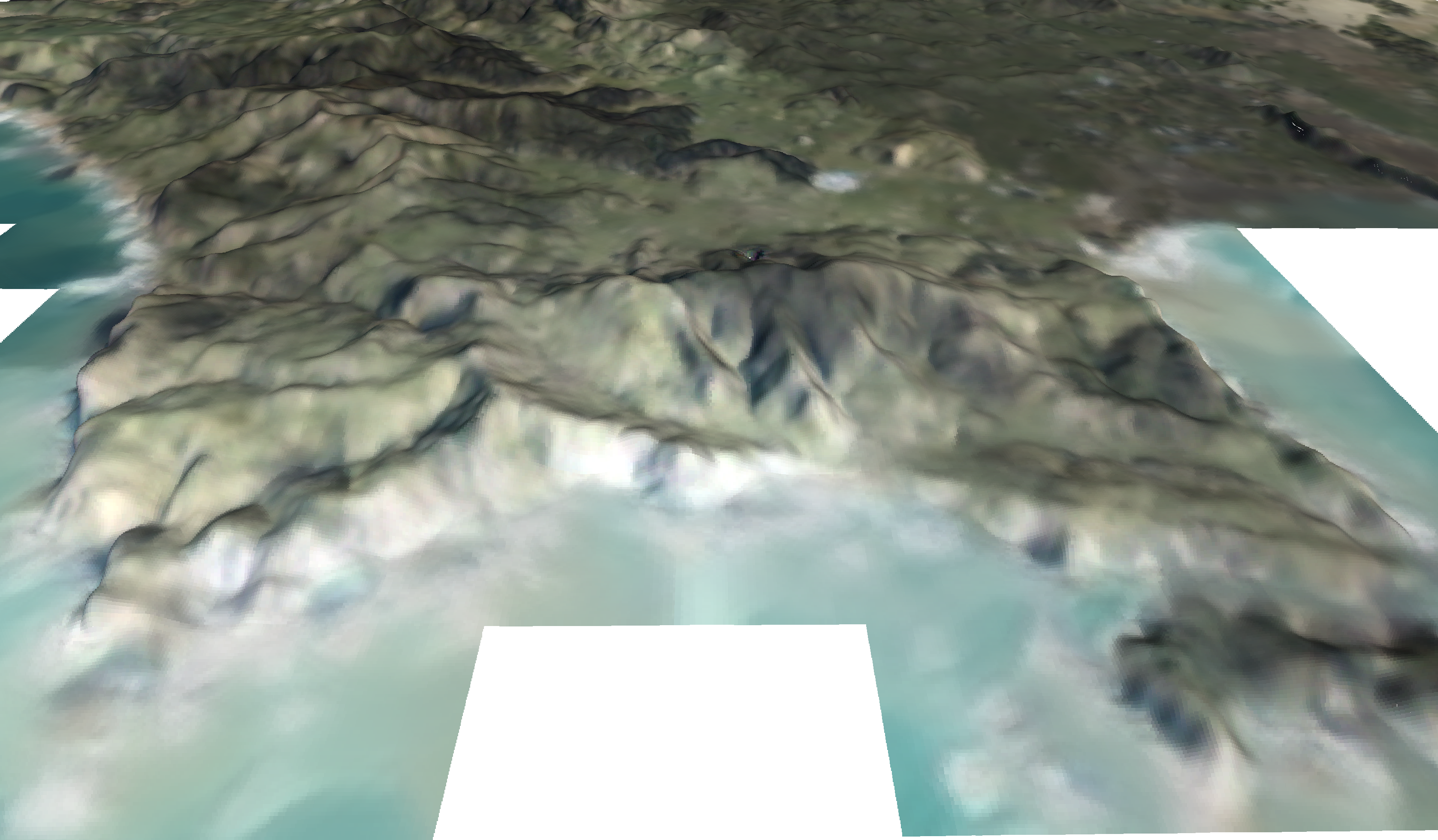}
    \end{subfigure}
    \hfill
    \begin{subfigure}[b]{0.30\textwidth}
        \centering
        \includegraphics[width=\textwidth]{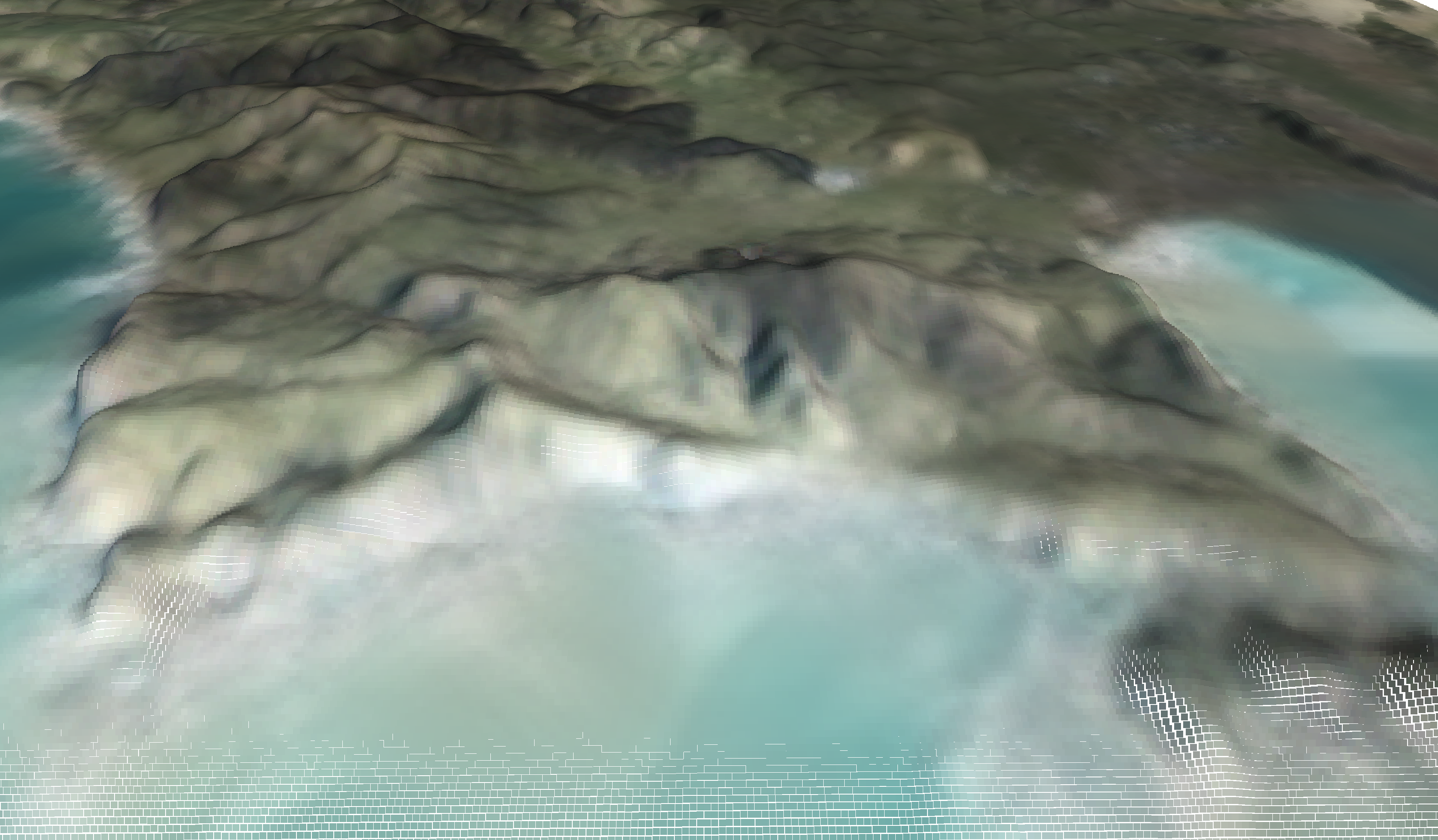}
    \end{subfigure}
    \hfill
    \begin{subfigure}[b]{0.30\textwidth}
        \centering
        \includegraphics[width=\textwidth]{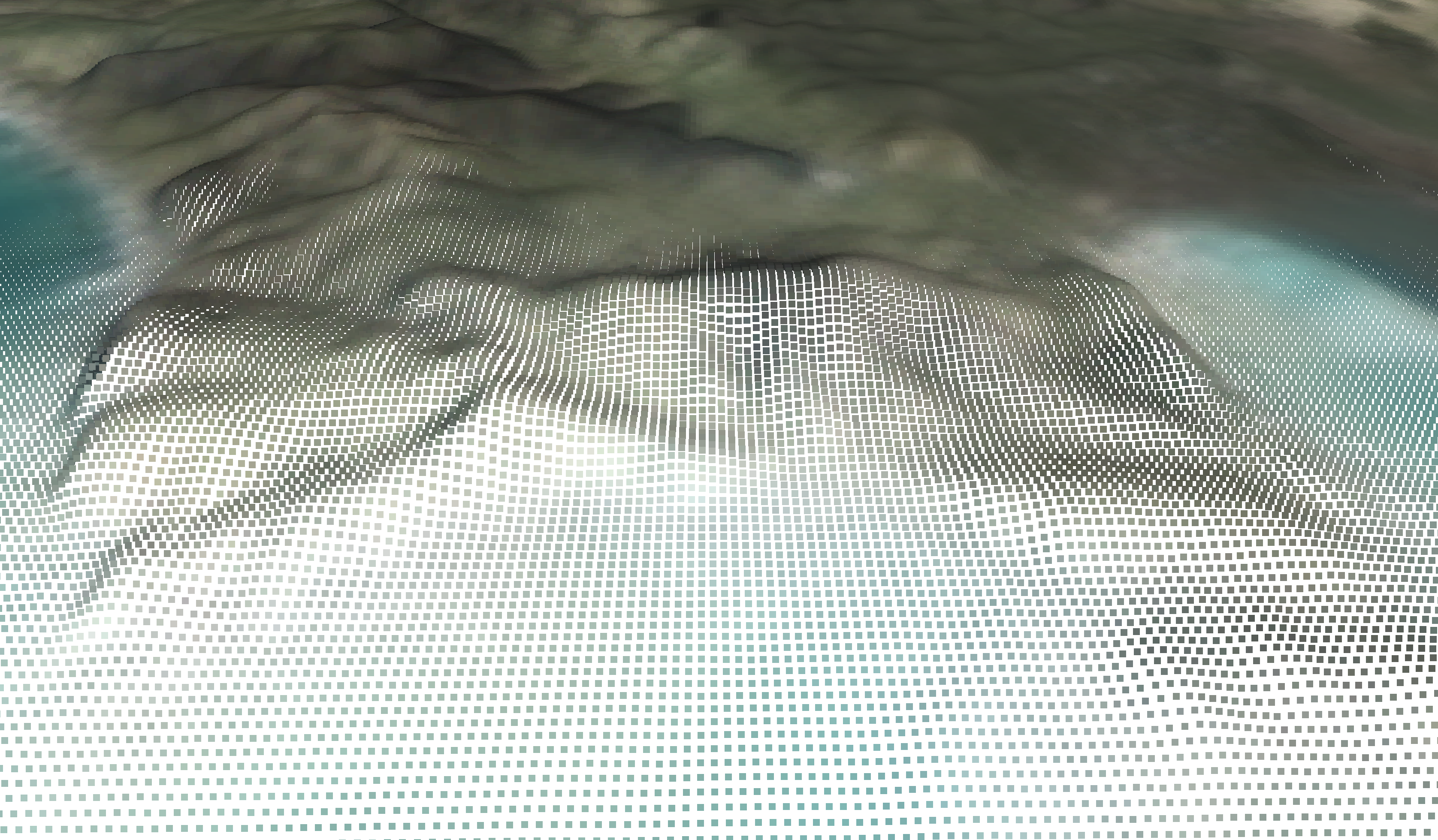}
    \end{subfigure}
    
    \caption{Reconstruction of \dataNzA with different resolutions. Upper: linear interpolation. Lower: \name. Left to right: 10 (default), 20, 40 meters per texel. }
    \label{fig:hm_res_gisborne}
\end{figure*}

\begin{figure*}[h]
\centering
\includegraphics[width=\linewidth]{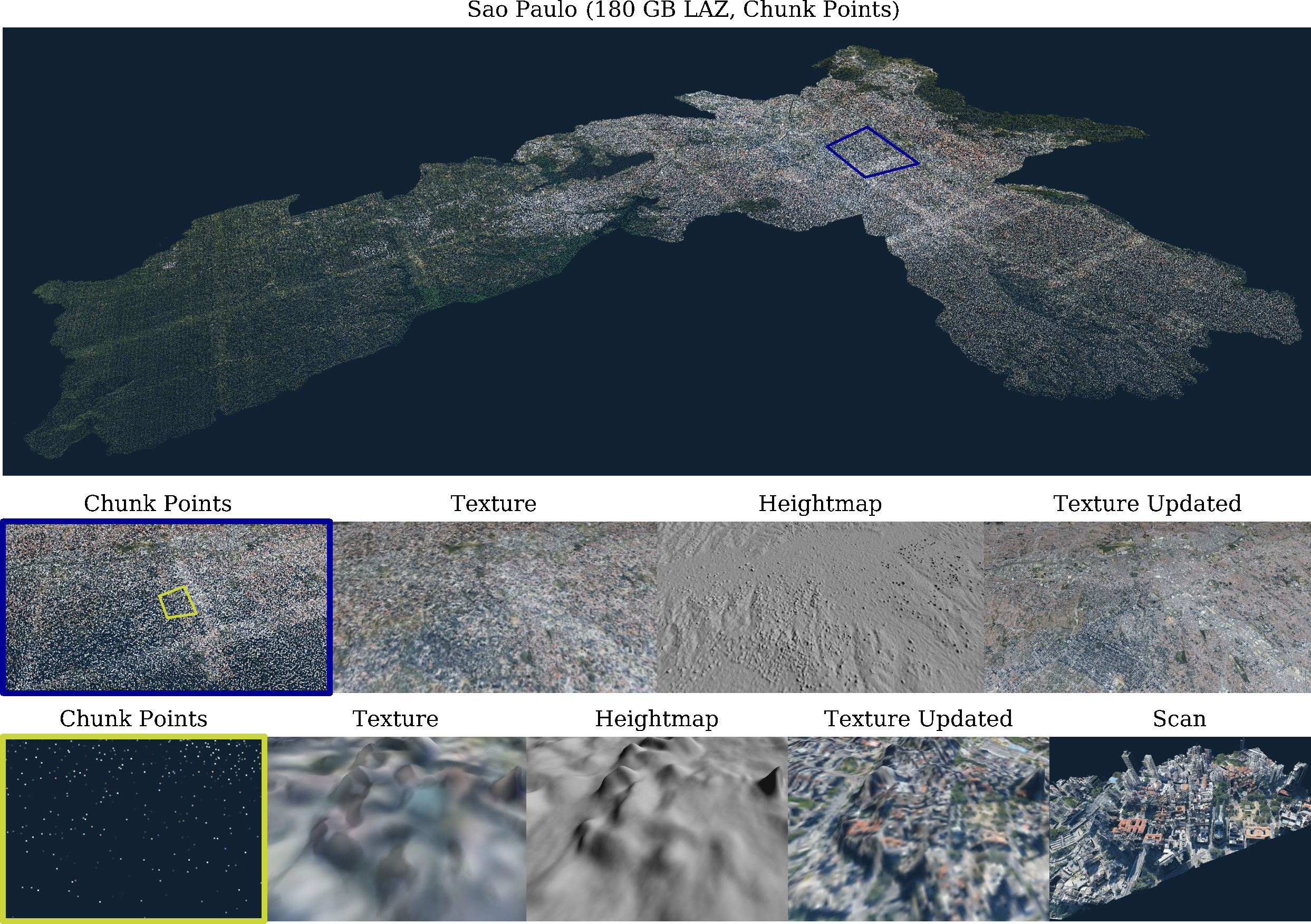}
\caption{\diff{Sao Paulo. Screenshots made of the urban center using \name. Source: Brazil Lidar Survey 2017 \url{https://portal.opentopography.org/lidarDataset?opentopoID=OTLAS.062020.31983.1}}}
\label{fig:sao_paulo}
\end{figure*}



\bibliographystyle{eg-alpha-doi}
\bibliography{main}



\end{document}